\documentclass[a4paper,10pt,twocolumn]{article}
\pdfoutput=1
\usepackage{graphicx}
\usepackage{xcolor}
\usepackage{url}
\usepackage[colorlinks=true,linkcolor=blue,citecolor=blue]{hyperref}
\usepackage[expproduct=cdot]{siunitx} %expproduct=tighttimes
\usepackage{relsize} % for relative font sizes, useful in commands such as PCa, Wnt\
\usepackage{amsmath}
\usepackage{amssymb}
\usepackage{mathptmx}\usepackage{upgreek}\newcommand{\betaup}{\upbeta} % text=times, with better 

\usepackage{combinedgraphics} % to rescale graphics and/or text parts independently in a combined graphics+text figure (gnuplot, inkscape)
\usepackage{enumitem} %\begin{enumerate}[itemsep=2pt,parsep=2pt,topsep=0pt,partopsep=0pt]

\usepackage[T1]{fontenc} % better to use type 1 fonts than bitmapped fonts

\usepackage{ctable}\newcommand{\thickmidrule}{\midrule[\heavyrulewidth]} % \toprule, \thickmidrule, \midrule, \cmidrule and \bottomrule to replace \hline in tables
\usepackage{tabularx} % For extensible columns. Alternative: tabulary. Calls \usepackage{array}
\usepackage{cite} % to display references [1, 2, 3] as [1-3] (without using natbib)

\definecolor{royalblue4}{HTML}{27408B}% royalblue4 in Emacs (see M-x list-colors-display)
\definecolor{red4}{HTML}{8B0000}% red4 in Emacs
\definecolor{green4}{HTML}{008b00} % green4 in Emacs

\usepackage{dblfloatfix} % to use modifiers [h] etc. in \begin{figure*} ... \end{figure*} environment
\usepackage{fixltx2e} % prevent starred-figures from appearing out of ascending order
\usepackage[font=footnotesize,labelfont=bf,labelsep=endash,margin=0pt]{caption} % have figure/table caption in smaller font with bf marker
\usepackage[title,header]{appendix}

\newlength{\myleftmargin} 
\newlength{\mytopmargin} 
\newlength{\myrightmargin} 
\newlength{\mybottommargin} 
\usepackage[runin]{abstract}

\let\paragraphold\paragraph
\renewcommand*{\paragraph}[1]{\paragraphold{#1.}} % automatically add a dot after argument of \paragraph
\newcommand{\keywords}[1]{\vspace{2mm}\noindent\textbf{Key words:} #1} % best to use within abstract
\newcommand{\pagewidetitle}[3] % to display #1=\maketitle, #2=abstract, and #3=title/author footnotes at the correct place
{%
    \twocolumn%
        [%
            \vskip-5mm%
            \begin{@twocolumnfalse}%
                #1%
                #2%
                \vspace{5mm}%
            \end{@twocolumnfalse}%
        ]%
        #3%
}

\newlength{\figurewidth}

\newcommand{\placeobj}[4][l]{%
    \makebox[0pt][#1]{\hspace{#2}\raisebox{-#3}[0pt][0pt]{#4}}%
}
\newcommand{\ie}{{\it i.e.}}

\newcommand{\eg}{{\it e.g.}}
\newcommand{\etal}{{\it et\ al.}}

\newcommand{\der}{\mathrm{d}}
\newcommand{\p}{\partial}
\newcommand{\pd}[2]{\frac{\partial #1}{\partial #2}}
\newcommand{\tpd}[2]{\tfrac{\partial #1}{\partial #2}}

\newcommand{\da}{\ensuremath{\text{day}}} %std?
\newcommand{\days}{\ensuremath{\text{days}}} %std?
\newcommand{\um}{\ensuremath{\micro\metre}}%  micro meter, from siunitx
\newcommand{\bmu}{\text{\textsmaller{BMU}}} % careful: if using \ensuremath, not typeset in boldface in titles
\newcommand{\Bmu}{\text{B\textsmaller{MU}}}
\newcommand{\ob}{\text{\textsmaller{OB}}}

\newcommand{\msc}{\textsmaller{MSC}} % bms cells
\newcommand{\obu}{\text{\textsmaller{OB}$_\text{u}$}} % alternative: \textsmaller{ob$_{\rm u}$} but then u is not changed to boldface in titles
\newcommand{\obumax}{\text{\textsmaller{OB}$_\text{u}^\text{max}$}}
\newcommand{\obp}{\text{\textsmaller{OB}$_\text{p}$}}

\newcommand{\oba}{\text{\textsmaller{OB}$_\text{a}$}}
\newcommand{\ocy}{\text{\textsmaller{OCY}}}
\newcommand{\blc}{\text{\textsmaller{LC}}}

\newcommand{\ocp}{\text{\textsmaller{OC}$_\text{p}$}}
\newcommand{\ocpmax}{\text{\textsmaller{OC}$_\text{p}^\text{max}$}}
\newcommand{\oca}{\text{\textsmaller{OC}$_\text{a}$}}

\newcommand{\tgfb}{\text{\textsmaller{TGF\textsmaller{$\betaup$}}}}
\newcommand{\Tgfb}{T\textsmaller{GF$\betaup$}}

\newcommand{\igf}{\text{\textsmaller{IGF}}}

\newcommand{\rank}{\text{\textsmaller{RANK}}}
\newcommand{\rankl}{\text{\textsmaller{RANKL}}}

\newcommand{\opg}{\text{\textsmaller{OPG}}}

\newcommand{\pth}{\text{\textsmaller{PTH}}}

\newcommand{\piact}{\ensuremath{\pi^\text{act}}} % first argument is 'receptor'
\newcommand{\pirep}{\ensuremath{\pi^\text{rep}}} % second argument is 'regulator'
\newcommand{\kform}{\text{$k_\text{form}$}} % alternative to \ensuremath
\newcommand{\kformdog}{\text{$k_\text{form}^\text{dog}$}} % alternative to \ensuremath
\newcommand{\kformtilde}{\ensuremath{\widetilde{k}_\text{form}}} % alternative to \ensuremath
\newcommand{\kres}{\text{$k_\text{res}$}}

\newcommand{\dobu}{\ensuremath{\mathcal{D}_\obu}}
\newcommand{\dobp}{\ensuremath{\mathcal{D}_\obp}}

\newcommand{\docp}{\ensuremath{\mathcal{D}_\ocp}}
\newcommand{\aoca}{\ensuremath{\mathcal{A}_\oca}}

\newcommand{\mar}{\text{MAR}}
\newcommand{\form}{\text{form}}
\newcommand{\RH}{\text{$R_\text{H}$}} % (Haversian canal radius)
\newcommand{\Rc}{\text{$R_\text{c}$}} % (Cement line radius)

\newcommand{\R}{\text{$R$}} % (BMU cavity radius)
\newcommand{\Sx}{\text{$S$}} % (BMU cavity cross-sectional area)
\newcommand{\Px}{\text{$P$}} % (BMU cavity cross-sectional perimeter)
\newcommand{\hOBa}{\text{$h_\oba$}} % (height of active osteoblasts - mixed biochemical/histomorph. notation)
\newcommand{\SOBa}{\text{$S_\oba$}} % (area of space occupied by osteoblasts as seen in cross-section - mixed biochemical/histomorphom. notation)

\newcommand{\bstrut}[1]{\rule[-#1]{0pt}{#1}} % bottom strut: to force space below the point where this is used: 
\newcommand{\tablestrut}{\rule{0pt}{3.8ex}} % top strut

\begin{document}
\title{\bf Bone refilling in cortical bone multicellular units: Insights into \\tetracycline double labelling from a computational model}

\author{Pascal R Buenzli,$^\text{1}$ Peter Pivonka, David W Smith}

\date{\small \vspace{-2mm}%$^\text{a}$
Faculty of Engineering, Computing \& Mathematics,
    \\The University of Western Australia, WA 6009, Australia\\\vskip 1mm \normalsize \today\vspace*{-5mm}}

\pagewidetitle{% Not strictly necessary. Remove this for standard LaTeX, or add \newcommand{\pagewidetitle}[3]{#1#2#3}
\maketitle
}{
\begin{abstract}
    Bone remodelling is carried out by `bone multicellular units' ($\bmu$s) in which active osteoclasts and active osteoblasts are spatially and temporally coupled. The refilling of new bone by osteoblasts towards the back of the $\bmu$ occurs at a rate that depends both on the number of osteoblasts and on their secretory activity. In cortical bone, a linear phenomenological relationship between matrix apposition rate and $\bmu$ cavity radius is found experimentally. How this relationship emerges from the combination of complex, nonlinear regulations of osteoblast number and secretory activity is unknown.

Here, we extend our previous mathematical model of cell development within a single cortical $\bmu$ to investigate how osteoblast number and osteoblast secretory activity vary along the $\bmu$'s closing cone. The mathematical model is based on biochemical coupling between osteoclasts and osteoblasts of various maturity, and includes the differentiation of osteoblasts into osteocytes and bone lining cells, as well as the influence of $\bmu$ cavity shrinkage on osteoblast development and activity. Matrix apposition rates predicted by the model are compared with data from tetracycline double labelling experiments. We find that the linear phenomenological relationship observed in these experiments between matrix apposition rate and $\bmu$ cavity radius holds for most of the refilling phase simulated by our model, but not near the start and end of refilling. This suggests that at a particular bone site undergoing remodelling, bone formation starts and ends rapidly, supporting the hypothesis that osteoblasts behave synchronously. Our model also suggests that part of the observed cross-sectional variability in tetracycline data may be due to different bone sites being refilled by $\bmu$s at different stages of their lifetime. The different stages of a $\bmu$'s lifetime (such as initiation stage, progression stage, and termination stage) depend on whether the cell populations within the $\bmu$ are still developing or have reached a quasi-steady state while travelling through bone. We find that due to their longer lifespan, active osteoblasts reach a quasi-steady distribution more slowly than active osteoclasts. We suggest that this fact may locally enlarge the Haversian canal diameter (due to a local lack of osteoblasts compared to osteoclasts) near the $\bmu$'s point of origin.

\end{abstract}
}{

\protect\footnotetext[1]{Corresponding\mbox{ }author.\mbox{ }Email\mbox{ }address:\\\texttt{<firstname>.<lastname>@uwa.edu.au}}
}% end of \pagewidetitle.

\section{Introduction}
Bone remodelling renews bone tissue in a spatially and temporally discrete fashion by means of `basic multicellular units' ($\bmu$s). In a $\bmu$, osteoclasts (bone-resorbing cells) create a resorption cavity called the `cutting cone' and osteoblasts (bone-forming cells) refill this cavity, forming a so-called `closing cone'~\cite{martin-burr-sharkey,parfitt-1994}. The action of osteoclasts and osteoblasts in a $\bmu$ is well coordinated such that bone formation closely follows bone resorption~\cite{martin-burr-sharkey,manolagas,jilka-2003,harada-rodan}. During remodelling, bone may be either gained or lost depending on (i) the final balance of bone turned over by a $\bmu$, \ie, the amount of bone refilled compared to the amount of bone resorbed; and (ii) the number of active $\bmu$s, due to the temporary $\bmu$ cavities that exist before refilling by the osteoblasts has completed~\cite{heaney,hernandez-beaupre-carter}. These mechanisms of gain or loss depend in particular on a fine regulation and timing of bone refilling in a $\bmu$. Under-refilling of the $\bmu$ cavity occurs for instance in advanced stages of osteoporosis~\cite{seeman,manolagas,jilka-2002,jilka-2003}, whilst over-refilling of the $\bmu$ cavity occurs in sclerosing bone disorders such as sclerosteosis, van Buchem disease, and high-bone-mass phenotype~\cite{vanBezooijen-etal}.\footnote{Over-refilling occurs mainly on trabecular and endosteal surfaces. However, a net bone gain is possible in cortical $\bmu$s that follow a pre-existing Haversian canal while reducing the canal's size (`type II osteon')~\cite{parfitt-in-recker,robling-stout}.} How bone refilling is controlled in a $\bmu$ remains poorly characterised. In particular the relative contributions of osteoblast number and osteoblast secretory activity at different phases of the refilling process are unknown. In this paper we develop a computational model of a $\bmu$ to investigate these contributions. The model will focus on a perfectly cylindrical geometry. Experimental data on the refilling dynamics in cortical bone are usually deduced from regular, cylindrical $\bmu$s.  However, general considerations on the refilling dynamics of $\bmu$s (Section~\ref{sec:bone-refilling-dynamics}) will be presented in an arbitrary geometry to elucidate how experimental measurements of osteoblast secretory activity could be deduced from irregular $\bmu$s (see also the Conclusions).

The matrix apposition rate at the back of a cortical $\bmu$ (\ie, the rate of change of the $\bmu$ cavity radius ($\R$)) depends both on the surface density of active osteoblasts $\rho_\oba$ and on their secretory activity $\kform$ (\ie, the volume of osteoid secreted per osteoblast per unit time) (see Section~\ref{sec:bone-refilling-dynamics}):
\begin{align}\label{mar-vs-kform-rhooba}
    \tpd{}{t}\R = -\kform\ \rho_\oba.
\end{align}
Experimentally, matrix apposition rates are estimated by dynamic histomorphometry techniques such as tetracycline double labelling~\cite{frost-1969,eriksen-histomorphometry,lee-1964,manson-waters,ilnicki-frost-etal,hood-etal,martin-dannucci-hood,metz-etal}. In cortical bone, such double labelling experiments reveal that matrix apposition rate and $\bmu$ cavity radius are linearly related~\cite{lee-1964,manson-waters,hood-etal,martin-dannucci-hood}:
\begin{align}\label{mar-vs-r}
    \tpd{}{t}\R = -C\ \R.
\end{align}
How this linear phenomenological relationship arises from underlying biological mechanisms in Eq.~\eqref{mar-vs-kform-rhooba} is still uncertain. Both the density of osteoblasts $\rho_\oba$ and their secretory activity $\kform$ vary as refilling proceeds~\cite{marotti-etal-1976}. The surface density of osteoblasts in the $\bmu$ $\rho_\oba$ depends on complex coupling to osteoclasts~\cite{parfitt-1994,manolagas,jilka-2003,harada-rodan}, bone surface availability, embedment in the bone matrix (transition to an osteocytic phenotype)~\cite{franzOdendaal-hall-witten}, and apoptosis~\cite{parfitt-1994,manolagas,jilka-2003}. The rate of osteoid secretion by a single osteoblast $\kform$ depends on phenotype (transition to a bone lining cell), the cell's protoplasmic volume~\cite{marotti-etal-1976,zambonin-zallone-1977,volpi-etal}, and may be regulated by the availability of nutrients, signalling by local regulatory molecules and hormones~\cite{manolagas,jilka-2003,harada-rodan}, the proximity of the blood vessel~\cite{martin-burr-sharkey}, and/or the local geometry of the bone substrate~\cite{qiu-parfitt-etal-2010,rumpler-fratzl-etal,martin-2000}.

Few works have investigated the number of osteoblasts and their level of secretory activity in a $\bmu$ \emph{in vivo}~\cite{jones,marotti-etal-1976,volpi-etal}, as also observed in~\cite{qiu-parfitt-etal-2010}. In Refs~\cite{jones,marotti-etal-1976,volpi-etal}, the secretory activity of osteoblasts was deduced  from measurements of cell surface density and matrix apposition rate, and use of Eq.~\eqref{mar-vs-kform-rhooba}. Marotti~\etal~\cite{marotti-etal-1976} performed serial sections of cortical $\bmu$s to do so, but a quantitative analysis is only presented for three cross-sectional slices.

The comparison of Eq.~\eqref{mar-vs-kform-rhooba} and Eq.~\eqref{mar-vs-r} emphasises the implicit influence of the $\bmu$ cavity radius $\R$ on the population of osteoblasts and/or on their secretory activity. Thus, changes in local bone geometry and changes in osteoblasts are interdependent. Influences of the local bone geometry may appear through signalling (\eg\ by osteocytes~\cite{marotti-2000,martin-2000}), and through the availability of bone surface to active osteoblasts. The experimental assessment of geometrical influences on bone formation \emph{in vivo} is challenging due the difficulty of controlling the geometry, but substrate geometry was shown to play an important role in an \emph{in vitro} system~\cite{rumpler-fratzl-etal}, and it could be responsible for the difference between matrix apposition rates in cortical and in trabecular bone~\cite{martin-2000}.

In this paper, we develop a mathematical model of a cortical $\bmu$ with the aim to understand (i) how osteoblast population size and osteoblast secretory activity vary with the various stages of the refilling process in a $\bmu$; and (ii) how a linear relationship between matrix apposition rate and $\bmu$ cavity radius, Eq.~\eqref{mar-vs-r}, arises from Eq.~\eqref{mar-vs-kform-rhooba} while accounting for complex, nonlinear biochemical and geometric regulatory mechanisms of osteoblast development. Specifically, the implicit dependence of $\rho_\oba$ upon $\R$ will be investigated by modelling at the cellular level the influence of the evolving cavity radius on the development of osteoblasts. The implicit dependence of $\kform$ upon $\R$ will be extrapolated from experimental data from Ref.~\cite{marotti-etal-1976}.

Determining how osteoblast number and osteoblast secretory activity vary along a $\bmu$ is critical to fully understand how bone refilling is regulated in a $\bmu$. This has important implications for several disorders of bone remodelling and their treatment, in particular for osteoporosis~\cite{seeman}. Geometric influences on bone cell development are also important to determine in order to assess the true contribution of biochemical regulatory systems. Experimental challenges associated with the study of geometric influences \emph{in vivo}  emphasise the need for accurate computational models that account for these geometric influences.

Mathematical and computational models of cell populations in bone remodelling are improving as a tool to assess the complexity of the spatio-temporal dynamics of cell--cell and cell--bone interactions. In recent years, several teams of researchers have developed such mathematical models, either focused on the tissue scale~\cite{komarova-etal,pivonka-etal1,vanOers-etal-2011,buenzli-etal-anabolic,pivonka-buenzli-etal-specific-surface,buenzli-etal-trabecularisation} or on the scale of a single $\bmu$~\cite{polig-jee,ryser-etal1,ryser-etal2,vanOers-etal-2008,ji-etal,buenzli-etal-moving-bmu,buenzli-etal-oca-resorption}. To our knowledge, the only previous mathematical model investigating the refilling process of a $\bmu$ in some detail is that of Ref.~\cite{polig-jee}. However, in Ref.~\cite{polig-jee}, the resorption process and the start of the refilling process are not modelled. The refilling process starts instead with an assumed initial population of active osteoblasts.

The model we develop here is based on biochemical regulation of osteoclast and osteoblast development across various maturity stages. It extends our previous model~\cite{buenzli-etal-moving-bmu} by adding important influences on the development of osteoblasts in the $\bmu$, namely, a geometrical effect of cavity shrinkage on the density of osteoblasts, an osteoblast-to-osteocyte transition (that likewise depends on the geometry), and an osteoblast-to-bone lining cell transition at the end of the refilling process. Accounting for these additional influences enables us to retrieve a combined evolution of osteoblast density and $\bmu$ cavity radius compatible with the linear decrease of matrix apposition rate with $\R$ observed from tetracycline experiments, Eq.~\eqref{mar-vs-r}. The direct comparison of our model with tetracycline data enables us to emphasise limitations of such data. Our model also provides insights into how the $\bmu$ evolves from its initiation stage (early-life) to its progression stage (quasi-steady state, mid-life).
Finally, in contrast to our previous model~\cite{buenzli-etal-moving-bmu}, the present model is calibrated for osteoclast, osteoblast, and precursor cell numbers in the $\bmu$~\cite{jaworski-duck-sekaly,parfitt-1994}, and for osteoblast surface densities at three radii of the $\bmu$ cavity~\cite{marotti-etal-1976}.

\section{Methods}\label{sec:methods}
We first provide a derivation of Eq.~\eqref{mar-vs-kform-rhooba}, generalised to $\bmu$s of any shape (Section~\ref{sec:bone-refilling-dynamics}). This derivation does not rely on a specific model of osteoblast population and activity. We then present our model of osteoclast and osteoblast development within the $\bmu$ (Section~\ref{sec:osteoblast-model}), and apply it to Eq.~\eqref{mar-vs-kform-rhooba} to determine matrix apposition rates in the closing cone of the $\bmu$.

\subsection{Bone refilling dynamics}\label{sec:bone-refilling-dynamics}
\begin{figure*}[!tbp]
    \centering
    \placeobj{0mm}{-56mm}{(a)}%
    \placeobj{110mm}{-56mm}{(b)}%
    \includegraphics[width=\textwidth]{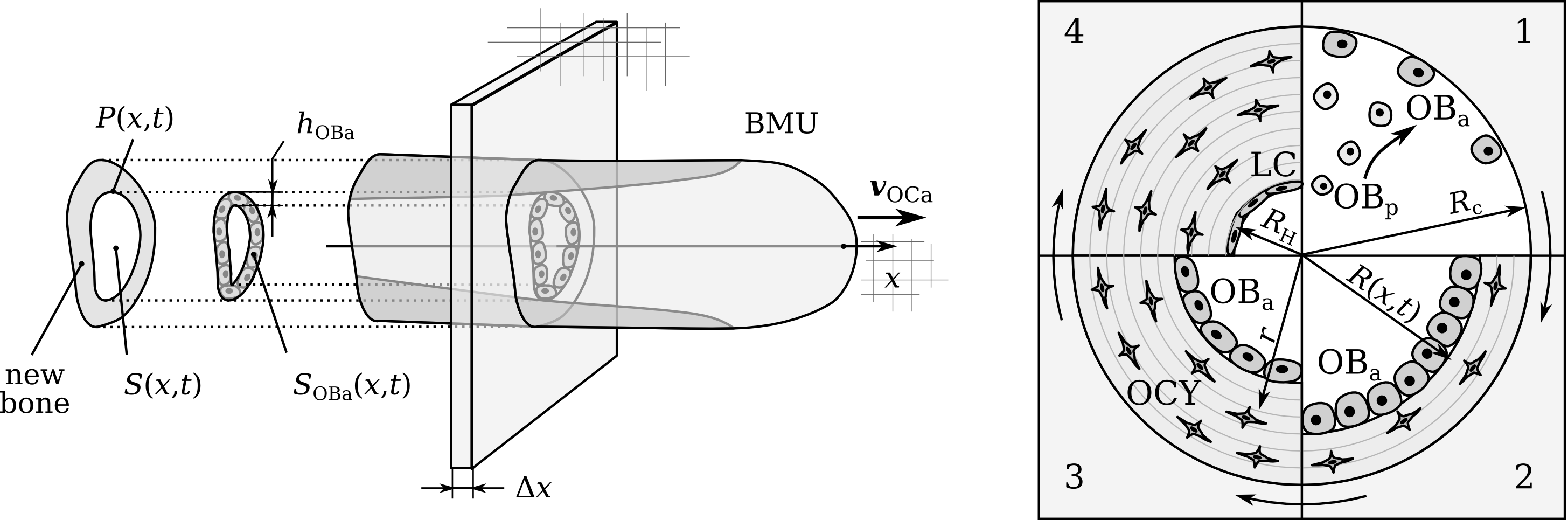}
    \caption{(a) The slice of bone of thickness $\Delta x$ at the fixed position $x$ reveals a transverse cross-section of the $\bmu$ (which itself progresses along the $x$ axis at speed $v_\oca$). The section exibits the ring-like cross-sectional area of newly-formed bone, the cross-sectional area of the cavity $\Sx(x,t)$, the cross-sectional perimeter of the cavity $\Px(x,t)$, and the ring-like cross-sectional area containing active osteoblasts $\SOBa(x,t)$, of width $\hOBa$. (b) Four stages of the formation phase of the $\bmu$ seen in transverse cross-section at the fixed location $x$ in bone as the $\bmu$ progresses forward along $x$: 1. Onset of formation: differentiation of pre-osteoblasts ($\obp$) into active osteoblasts ($\oba$); cavity radius equal to the cement line radius $\Rc$; 2.  Early stage of formation, showing that some osteoblasts have become osteocytes ($\ocy$) embedded in the bone matrix; cavity radius $\R(x,t)$; 3. Mid stage of formation; $r$ denotes a radial coordinate; 4. End of formation: the active osteoblasts have become bone lining cells ($\blc$) and the cavity radius is the Harversian canal radius $\RH$.}
    \label{fig:bmu-geom}
\end{figure*}
We consider a thin slice of bone exhibiting the refilling process in the closing cone of a $\bmu$ (Figure~\ref{fig:bmu-geom}a). The slice corresponds to a local region of interest and is assumed to be (i) transverse to the $\bmu$'s longitudinal axis ($x$); (ii) of small thickness $\Delta x$; and (iii) at a fixed position~$x$ in bone (but the state of the $\bmu$ at various positions $x$ will be investigated).

As the $\bmu$ progresses forward along $x$, gradually moving away from the slice, the $\bmu$ cavity (Cv) seen in the slice refills with time due to the local production of osteoid by active osteoblasts (Figure~\ref{fig:bmu-geom}b). The total volume of osteoid produced per unit time in the slice is $\kform(x,t)\Delta N_\oba(x,t)$, where $\kform(x,t)~[\um^3/\da]$ is the volume of osteoid secreted per osteoblast per unit time, and $\Delta N_\oba(x,t)$ is the number of active osteoblasts in the slice. The production of osteoid reduces the cross-sectional area of the $\bmu$ cavity $\Sx(x,t)$ (Figure~\ref{fig:bmu-geom}a) according to:
\begin{align}\label{refilling-eq-number}
    \tfrac{\p}{\p t}\Sx(x,t)\,\Delta x = - \kform(x,t)\ \Delta N_\oba(x,t).
\end{align}
Dividing Eq.~\eqref{refilling-eq-number} by $\Delta x$, this can be expressed as:
\begin{align}\label{refilling-eq-surface}
    \tfrac{\p}{\p t} \Sx(x,t) = -\kform(x,t)\ \rho_\oba(x,t)\ \Px(x,t),
\end{align}
where
\begin{align}\label{oba-surface-density}
    \rho_\oba(x,t) = \frac{\Delta N_\oba(x,t)}{\Delta x\, \Px(x,t)}
\end{align}
is the local surface density of active osteoblasts and $\Px(x,t)$ is the cross-sectional perimeter of the $\bmu$ cavity (Figure~\ref{fig:bmu-geom}a). For a perfectly cylindrical cortical $\bmu$ of cavity radius $\R(x,t)$, $\Px(x,t) = 2\pi \R(x,t)$, $\Sx(x,t)=\pi \R(x,t)^2$, and so $\tfrac{\p}{\p t}\Sx = 2\pi \R \tfrac{\p}{\p t}\R$. In this case, Eq.~\eqref{refilling-eq-surface} specialises to Eq.~\eqref{mar-vs-kform-rhooba}:
\begin{align}\label{refilling-eq-radius}
    \tfrac{\p}{\p t} \R(x,t) = -\kform(x,t)\ \rho_\oba(x,t).
\end{align}

The refilling dynamics\eqref{refilling-eq-surface} and \eqref{refilling-eq-radius} are based on geometrical considerations only. These expressions do not depend on a specific model of osteoblast surface density $\rho_\oba$ or osteoid secretion rate $\kform$. In this paper, we will restrict to the refilling dynamics of cylindrical $\bmu$s, Eq.~\eqref{refilling-eq-radius}.

\subsection{Computational model for osteoblast distribution in the $\bmu$}\label{sec:osteoblast-model}
The mathematical model presented in this paper is based on our previous mathematical model of osteoclast and osteoblast development in a cortical $\bmu$ in one spatial dimension (the $\bmu$ longitudinal axis)~\cite{buenzli-etal-moving-bmu}. Biochemical coupling between osteoclasts and osteoblasts and migration properties of these cells are implemented via the material balance equation expressed for each cell type and signalling molecule considered~\cite{buenzli-etal-moving-bmu}. These equations integrate source and sink terms due to cell differentiation and cell apoptosis, receptor--ligand binding reactions, protein expression by cells, micro-environmental protein degradation, as well as transport terms due to cell migration and protein diffusion. The core assumptions of the model are summarised below with the network of biochemical regulation illustrated in Figure~\ref{fig:bmu-model}. Further details are presented in Appendix~\ref{appx:model}, including all the governing equations and parameters of the model.
\begin{figure*}[!htbp]
    \centering\includegraphics[width=0.8\textwidth]{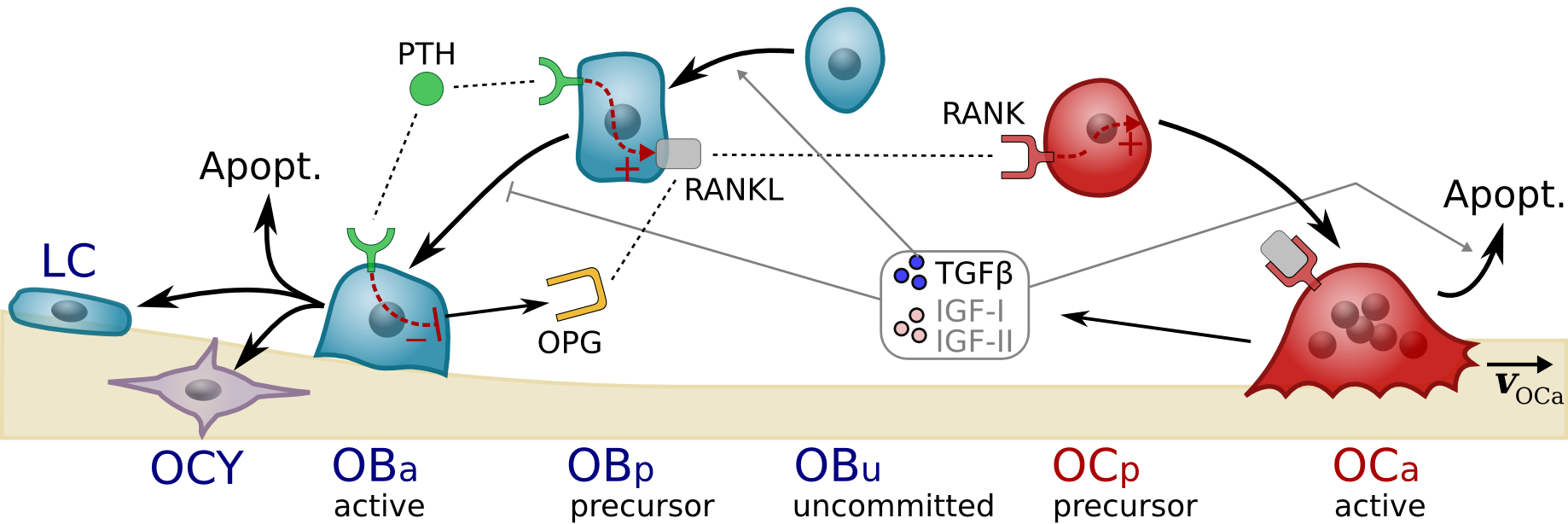}
    \caption{Biochemical regulation of osteoclast and osteoblast development in the $\bmu$ used in the model.}\label{fig:bmu-model}
\end{figure*}

\paragraph{$\Bmu$ cavity}
The $\bmu$ cavity is assumed to be perfectly cylindrical (rotation-symmetric with respect to the $x$ axis). Only the refilling of the cavity is considered in this paper. The cavity radius $R(x,t)$ is assumed to decrease from an initial value equal to the cement line radius $R_c$ up to the Haversian canal radius $R_H$ (at most) due to matrix deposition by osteoblasts (Eq.~\eqref{refilling-eq-radius}).\footnote{When $R=R_H$, osteoblast secretory activity is assumed to stop (see the \emph{Osteoblast} section below), but refilling may stop before $R$ reaches $R_H$ if the population of osteoblasts goes extinct.} The two parameters $R_c\approx 100~\um$ and $R_H\approx 20~\um$ correspond to human average values (see Table~\ref{table:parameters} in Appendix~\ref{appx:model}). No cross-sectional variability is assumed for $R_c$ and $R_H$ in this paper. The model results will be compared with tetracycline data (Figure~\ref{fig:mar}) that have been first normalised adequately to remove the influence of cross-sectional variability in cement line and Haversian canal radii \cite{metz-etal}, and that have been then rescaled to the values $R_c$ and $R_H$ assumed here.

\paragraph{Osteoclasts}
Two stages of osteoclast development are modelled: pre-osteoclasts ($\ocp$) and active osteoclasts ($\oca$) (Figure~\ref{fig:bmu-model}): 
\begin{itemize}[itemsep=2pt,parsep=1pt,topsep=3pt,partopsep=0pt]
    \item $\ocp$s are distributed around the tip of a blood vessel growing at a rate commensurate with the $\bmu$ progression ($20$--$40~\um/\da$)~\cite{martin-burr-sharkey,parfitt-1994,parfitt-1998,parfitt-in-recker}, taken to be $30~\um/\da$;
    \item $\ocp$s differentiate into $\oca$s at a rate $\docp$ accelerated by activation of the receptor--activator of nuclear factor $\kappa$B ($\rank$) by the ligand $\rankl$~\cite{roodman,martinTJ};
    \item $\oca$s undergo apoptosis at a rate $\aoca$ accelerated by transforming growth factor~$\beta$ (\tgfb)~\cite{roodman,pivonka-etal1};
    \item $\oca$s progress forward along the $x$ axis due to bone resorption, at a rate $v_\oca=30~\um/\da$~\cite{martin-burr-sharkey,parfitt-1994,parfitt-in-recker}.
\end{itemize}%
As a result, the material balance equation for the local volumetric density of active osteoclasts $\oca(x,t)~[\mm^{-3}]$ is~\cite{buenzli-etal-moving-bmu}:
\begin{align}\label{oca}
    \tpd{}{t}\oca = \docp(\rankl)\,\ocp - \aoca(\tgfb)\,\oca-\tpd{}{x}(\oca\, v_\oca).
\end{align}

\paragraph{Osteoblasts}
Three stages of osteoblast development are modelled: uncommitted osteoblast progenitors (such as mesenchymal stem cells) ($\obu$), pre-osteoblasts ($\obp$) and active osteoblasts ($\oba$). Osteoblasts may also become osteocytes ($\ocy$) or bone lining cells ($\blc$) (Figure~\ref{fig:bmu-model}):
\begin{itemize}[itemsep=2pt,parsep=1pt,topsep=3pt,partopsep=0pt]
    \item $\obu$s are distributed around the tip of the blood vessel growing along $x$ at $30~\um/\da$, similarly to $\ocp$s~\cite{martin-burr-sharkey,jaworski-hooper,parfitt-1994,parfitt-1998};
    \item $\obu$s differentiate into $\obp$s at a rate $\dobu$ accelerated by $\tgfb$~\cite{parfitt-1994,harada-rodan,iqbal-sun-zaidi,tang-etal};
    \item $\obp$s differentiate into $\oba$s at a rate $\dobp$ inhibited by $\tgfb$~\cite{pivonka-etal1,buenzli-etal-moving-bmu};
    \item $\obp$s and $\oba$s are stationary with respect to bone~\cite{parfitt-1994,martin-burr-sharkey,buenzli-etal-moving-bmu,pazzaglia-etal-2010,pazzaglia-etal-2012}.
\end{itemize}%
The following features of osteoblast development are added in the present work compared to Ref.~\cite{buenzli-etal-moving-bmu} (see Appendix~\ref{appx:model} for more details):
\begin{itemize}[itemsep=2pt,parsep=1pt,topsep=3pt,partopsep=0pt]
    \item \emph{Secretory activity.} The secretory activity of $\oba$s depends on the current cavity radius:
    \begin{align}
            \kform(x,t)=\kformtilde\big(\R(x,t)\big)\quad [\um^3/\da]\label{kform}
    \end{align} 
The function $\kformtilde$ elucidating this dependence is extrapolated from measurements from \cite{marotti-etal-1976}. The activity of osteoblasts is thereby assumed to vary continuously (\eg, no work--rest cycles) \cite{parfitt-in-recker,martin-burr-sharkey,pazzaglia-etal-2012-lamellar};
    \item \emph{Osteoblast-to-osteocyte transition.} $\oba$s transition to $\ocy$s at a rate $\sigma_\ocy^\text{prod.}$ that depends on the density of $\oba$s and on $\R$ \cite{franzOdendaal-hall-witten,dallas-bonewald,polig-jee}, such that a uniform density of $\ocy$s of $\num{20000}/\mm^3$ is produced \cite{parfitt-in-recker};
    \item \emph{Osteoblast-to-bone lining cell transition.} $\oba$s transition to $\blc$ when $\R$ reaches a target Haversian canal radius $\RH$. This is modelled by a sudden, but continuous drop in $\oba$s' secretory activity $\kformtilde(\R)$ near $\RH$ \cite{martin-burr-sharkey} (see Figure~\ref{fig:kform} in Appendix~\ref{appx:model});
    \item \emph{Osteoblast apoptosis.} $\oba$s undergo apoptosis at a constant rate $A_\oba$ calibrated so as to eliminate osteoblasts `in surplus' \cite{martin-burr-sharkey,parfitt-1994};
    \item \emph{Geometric influence of $\bmu$ cavity shrinkage.} The shrinking $\bmu$ cavity confines $\oba$s into a smaller space, which increases the density of $\oba$s at a rate $G$ that depends on $\R$ \cite{polig-jee,pazzaglia-etal-2011}.
\end{itemize}
The total change in the density of $\oba$s is a superposition of all of the influences listed above, which may compensate each other. In particular, the density of $\oba$s may still decrease during $\bmu$ cavity shrinkage if there is a concurrent decrease in the number of $\oba$s in the cavity. As a result of the above assumptions, the material balance equations for the local volumetric densities of pre-osteoblasts $\obp(x,t)$ and active osteoblasts $\oba(x,t)$ are:
\begin{align}
    &\tpd{}{t}\obp = \dobu(\tgfb)\,\obu - \dobp(\tgfb)\,\obp \label{obp}
    \\&\tpd{}{t}\oba = \dobp(\tgfb)\,\obp - A_\oba\,\oba + G\, \oba - \sigma_\ocy^\text{prod.} \label{oba-balance-avg-implicit}
\end{align}

The last three terms of Eq.~\eqref{oba-balance-avg-implicit} replace the constant elimination rate of $\oba$s used in Ref.~\cite{buenzli-etal-moving-bmu}, which was previously modelling all the depletion pathways of the pool of active osteoblasts (\ie, osteoblast-to-osteocyte, osteoblast-to-bone lining cell, and apoptosis). The precise form of these new terms (in particular, their dependence on the cavity radius $\R(x,t)$) is presented in Appendix~\ref{appx:model}.

\paragraph{Signalling molecules and binding reactions}
    The biochemical coupling between osteoclasts and osteoblasts is mediated in the model by $\rankl$ and $\tgfb$, as in Refs~\cite{pivonka-etal1,buenzli-etal-moving-bmu} ($\tgfb$ may also represent other regulatory factors stored in the bone matrix, such as insulin-like growth factors ($\igf$), see Figure~\ref{fig:bmu-model}). $\rankl$ is influenced by osteoprotegerin ($\opg$) and parathyroid hormone ($\pth$). All of $\rankl$, $\opg$ and $\tgfb$ are driven by cellular actions (Figure~\ref{fig:bmu-model}):
\begin{itemize}[itemsep=2pt,parsep=1pt,topsep=3pt,partopsep=0pt]
    \item $\rankl$ is expressed on the membrane of $\obp$s; $\opg$ is expressed by $\oba$s; $\rank$ is expressed on $\ocp$s~\cite{gori-etal,thomas-etal,buenzli-etal-moving-bmu};
    \item $\rankl$ binds both $\rank$ and the decoy receptor $\opg$ competitively ($\rank$ signalling on $\ocp$ is thereby diminished in presence of $\opg$)~\cite{roodman,martinTJ};
    \item $\pth$ promotes the expression of $\rankl$ and inhibits the expression of $\opg$~\cite{ma-martin-etal}. In this work, a constant level of systemic $\pth$ is assumed, uniformly distributed along the $\bmu$~\cite{buenzli-etal-moving-bmu}.
    \item $\tgfb$ is stored in the bone matrix at a constant concentration $n^\text{bone}_\tgfb$; it is released in the micro-environment in proportion to bone resorption by $\oca$s~\cite{roodman,iqbal-sun-zaidi,tang-etal};
    \item All of the signalling molecules $\rankl$, $\opg$, and $\tgfb$ undergo micro-environmental degradation at a constant rate. A constant number of $\rank$ per $\ocp$ is assumed.
\end{itemize}
Binding reactions between receptors and ligands are considered through mass action kinetics with reaction rates proportional to population sizes. The acceleration or inhibition of a cell behaviour by a ligand, such as differentiation rate, apoptosis rate, and protein production rate, is assumed to occur in proportion to the receptor occupancy on the cell (see Appendix~\ref{appx:model} and Ref.~\cite{buenzli-etal-moving-bmu} for further details).

\paragraph{Surface density and volumetric density of active osteoblasts}
Biochemical processes of cell development are best described in terms of volumetric concentrations of signalling molecules and volumetric density of cells. (To align with common practice, we use the terminology `concentration' for signalling molecules and `density' for cells, even though both terminologies refer to the same units, \ie, number per unit volume.) Indeed, biochemical reaction rates depend on the probability of encounter of the interacting biochemical compounds. This probability depends in turn on how closely packed the compounds are, and so, on their local concentration or density (\ie, their number per unit volume)~\cite{vankampen}. 

The material-balance equations~\eqref{oca}--\eqref{oba-balance-avg-implicit}, which involve biochemical reactions, are therefore written in terms of volumetric densities. However, the refilling dynamics described by Eq.\eqref{refilling-eq-radius} involves the surface density $\rho_\oba$ of active osteoblasts. To relate the volumetric density $\oba$ to the surface density $\rho_\oba$, we use the fact that active osteoblasts in a $\bmu$ form a single layer of cells against the cavity walls~\cite{marotti-etal-1976}, of thickness $\hOBa$ (the height of an active osteoblast). In the cross section, active osteoblasts are thus confined to an annulus of cross-sectional area $\SOBa$  (Figure~\ref{fig:bmu-geom}a). Within this annulus, active osteoblasts are distributed fairly uniformly, and so $\SOBa$ constitutes an appropriate representative surface element to define the volumetric density of active osteoblasts~\cite[Chap.~XIV]{vankampen}, \ie:
\begin{align}\label{oba-def}
    \oba(x,t) = \frac{\Delta N_\oba(x,t)}{\Delta x\ \SOBa(x,t)} = \rho_\oba(x,t) \frac{\Px(x,t)}{\SOBa(x,t)},
\end{align}
where Eq.~\eqref{oba-surface-density} has been used for the second equality. In a cylindrical $\bmu$, $\Px(x,t)=2\pi \R(x,t)$ and
\begin{align}\label{soba}
    \SOBa(x,t) = \pi\big[\R(x,t)^2 - (\R(x,t)-\hOBa)^2\big].
\end{align}
The relation~\eqref{oba-def} between osteoblast surface density and volumetric density becomes:
\begin{align}\label{rhooba-oba}
    \rho_\oba(x,t) = \oba(x,t)\ \hOBa \left(1-\frac{\hOBa}{2\R(x,t)}\right).
\end{align}
Substituting Eq.~\eqref{rhooba-oba} into Eq.~\eqref{refilling-eq-radius} gives:
\begin{align}\label{refilling-eq-radius-oba}
    \tfrac{\p}{\p t} \R(x,\!t) = -\kform(x,\!t)\oba(x,\!t)\hOBa\!\left(\!\!1\!-\!\frac{\hOBa}{2\R(x,\!t)}\!\!\right)\!.
\end{align}
We note that in the above relationship, the geometric factor $\big(1-\frac{\hOBa}{2\R(x,t)}\big)$ accounts for the nonzero curvature $1/\R$ of the cavity (a flat surface has zero curvature, corresponding to the limiting case $1/\R\to 0$). However, both $\kform$ and $\oba$ may still depend on $\R$.

\paragraph{Calibration of the cell populations and validation of the model}
Quantitative data on the distribution of cells and their total number in cortical $\bmu$s are relatively sparse~\cite{jaworski-hooper,jaworski-duck-sekaly,marotti-etal-1976,parfitt-1994}. Some of the data useful to calibrate and validate our model are derived from animal models, such as cell surface densities and osteoid secretion rates at different $\bmu$ cavity radii (dogs)~\cite{marotti-etal-1976}, total cell numbers (dogs)~\cite{jaworski-duck-sekaly}, and tetracycline double labelling radii (sheep)~\cite{metz-etal}. Some of these data were rescaled to (\mbox{pseudo-})human values according to known (or suspected) cross-species differences~\cite{jowsey} (Appendix~\ref{appx:model}). We calibrated our model such that the numerical cell distribution profiles are in reasonable agreement with the following:
\begin{itemize}[itemsep=2pt,parsep=1pt,topsep=3pt,partopsep=0pt]
    \item The total number of precursor cells ($N_\obu, N_\ocp$) in steady state is about 20;
    \item The total number of active osteoclast nuclei ($N_\oca$) in steady state is about 100~\cite{jaworski-duck-sekaly}. (Here and in Ref.~\cite{buenzli-etal-moving-bmu}, $\oca$s represent mononucleated entities incorporated in a multinucleated active osteoclast. Multinucleated active osteoclasts are composed of about 10 nuclei~\cite{jaworski-duck-sekaly};)
    \item The surface density of active osteoblasts ($\rho_\oba$) in steady state coincides with data reported by Marotti\ \etal~\cite{marotti-etal-1976} at three different cavity radii (appropriately rescaled), whilst the total number of active osteoblasts ($N_\oba$) in steady state is in the range $2000$--$6000$~\cite{jaworski-duck-sekaly,parfitt-in-recker,polig-jee} (see also Discussion, Section~\ref{sec:discussion}).
\end{itemize}

To calibrate the model against the number of active osteoclast nuclei and precursor cells in the steady-state $\bmu$, the cell densities $\oca(x,t)$, $\ocp(x,t)$ and $\obu(x,t)$ were integrated over the $\bmu$ cavity volume at $t=150~\days$. Appropriate changes in model parameter values were determined to enable these cell numbers to match the values mentioned above (Appendix~\ref{appx:model}, Eqs~\eqref{scaling}).

To calibrate the model against the surface densities of active osteoblasts at three different cavity radii (Table~\ref{table:ob-density}) and against the total number of active osteoblasts, we modified (i) the rate of osteoblast apoptosis $A_\oba$; (ii) the model parameters related to the scaling factor $\alpha_\oba$ (Appendix~\ref{appx:model}); and (iii) the (canine-to-human) scaling factor $\alpha_\kform$ (Appendix~\ref{appx:model}). The apoptosis rate $A_\oba$ influences the spatial rate of decrease of the osteoblast distribution at the back of the $\bmu$~\cite{buenzli-etal-moving-bmu}. The factor $\alpha_\oba$ scales the value of the density at each location along the $\bmu$ uniformly. The scaling factor $\alpha_\kform$ enables to modify the total number of active osteoblasts in the steady state $\bmu$ (\eg, increasing $\alpha_\kform$ reduces $N_\oba$). The total number of pre-osteoblasts was obtained by integrating $\obp(x,t)$ over the cavity volume and the total number of active osteoblasts was obtained by integrating the surface density $\rho_\oba(x,t)$ over the cavity--bone surface in the steady state $\bmu$ (at $t=150~\days$).

The calibrated model was validated against three independent sets of experimental data: (i) the density of bone lining cells reached at the end of the formation phase; (ii) the length of the closing cone of the $\bmu$; (iii) tetracycling double labelling data (see Sections~\ref{sec:results} and~\ref{sec:discussion}).

\paragraph{Governing equations and numerical simulations}
Equations~\eqref{oca}--\eqref{oba-balance-avg-implicit} and Eq.~\eqref{refilling-eq-radius-oba} govern the dynamics of the cell populations in the $\bmu$ and of the $\bmu$ cavity refilling. These equations are interdependent and involve binding reactions for the concentrations of $\rankl$, $\opg$, and $\tgfb$. $\tgfb$ has a production rate that depends on osteoclast resorption, but the dynamics of the signalling molecules $\rankl$ and $\opg$ is fast compared to the characteristic times of cell behaviours~\cite{buenzli-etal-moving-bmu}. The balance equations of $\rankl$ and $\opg$ can be taken in a quasi-steady state, enabling the concentration of these signalling molecules to become algebraic expressions of the variables $\obp, \oba, \oca$, and $\tgfb$~\cite{buenzli-etal-moving-bmu}. 

The equations governing the evolution of the whole system are presented in Appendix~\ref{appx:model}, see Eqs~\eqref{governing-eqs-pdes}--\eqref{piact-pirep}. These equations together with Eq.~\eqref{refilling-eq-radius-oba} form a coupled system of five partial differential equations (PDEs). These PDEs were solved numerically to evolve the system for 150 days from an initial condition with a small, localised population of active osteoclast nuclei ($\oca$s) and given distributions of $\obu$s and $\ocp$s concentrated around the (growing) tip of the blood vessel of the $\bmu$, as in Ref.~\cite{buenzli-etal-moving-bmu}. The initial cavity radius was set to the cement line radius $\Rc$ as only the refilling of the cavity is considered in this paper. The point of origin of the $\bmu$ (\ie, the center of the initial distribution of active osteoclasts and precursor cells) was set at $x_0 = -4.85~\mm$. The PDEs were solved in a co-moving frame attached to the $\bmu$ with the following boundary conditions: the density of pre-osteoblasts and active osteoblasts was set to zero at the (moving) front of the $\bmu$; the density of active osteoclasts and the concentration of $\tgfb$ were set to zero at the back of the $\bmu$. The numerical algorithm used was that of the `method of lines' of \texttt{Mathematica}'s PDE solver `\texttt{NDSolve}' with a spatial discretisation of at least 2000 points, ensuring absolute and relative tolerances of $10^{-4}$~\cite{mathematica}.

\section{Results}\label{sec:results}

\begin{figure}[!tp]
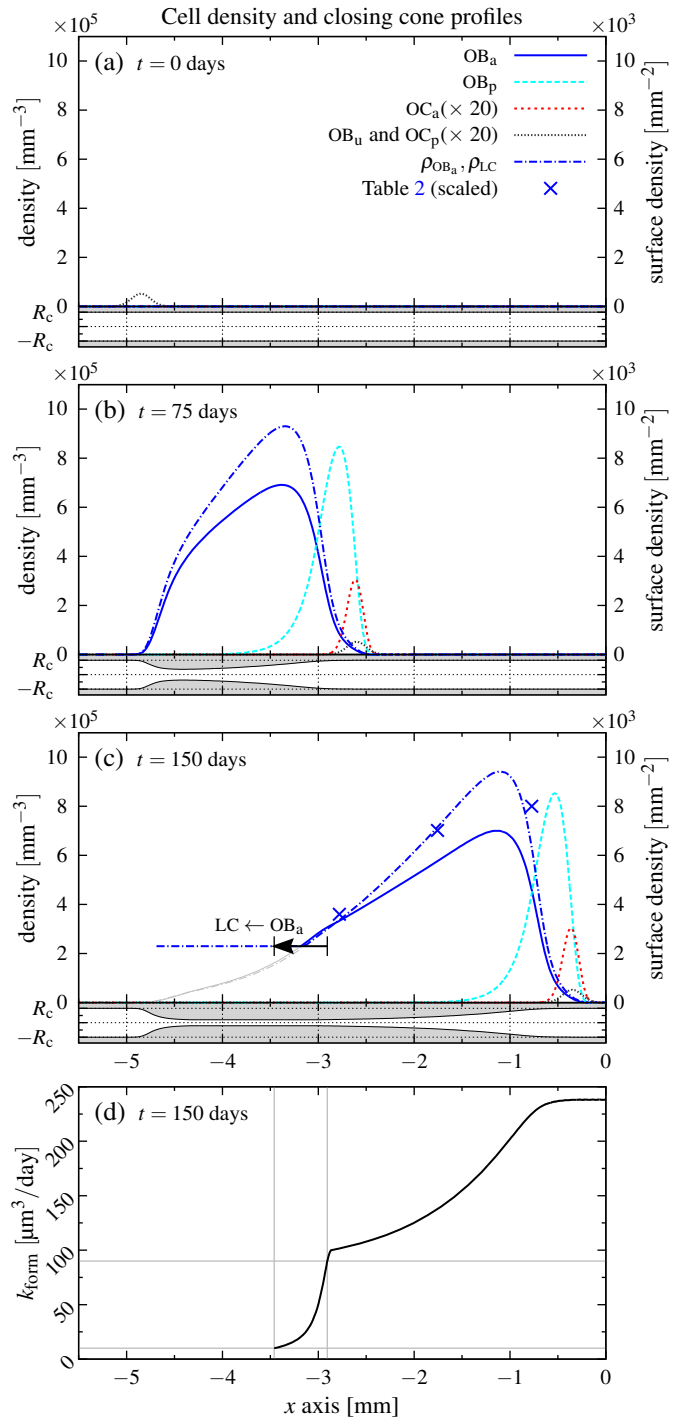

        \centering
        \makebox[\figurewidth]{\input{fig-profiles-0d}}%
        \vspace{-5mm}
        \makebox[\figurewidth]{\input{fig-profiles-75d}}%
        \vspace{-5mm}
        \makebox[\figurewidth]{\input{fig-profiles-150d}}%
        \vspace{-1.5mm}
        \makebox[\figurewidth]{\input{fig-kform-150d}}%
        \caption{(a)--(c) The profiles of cell densities along the $\bmu$'s length (top of each panel) and the corresponding shape of the closing cone (bottom of each panel) are shown at time (a) $t=0\ \days$ (initial condition), (b) $t=75\ \days$, and (c) $t=150\ \days$. To ensure that all cell densities are visible, $\obu$, $\ocp$, and $\oca$ were multiplied by a factor 20. The active osteoblast surface density $\rho_\oba$ and bone lining cell surface density $\rho_\blc$ are also shown (dot-dashed lines). In (b), individual osteoid secretion rate $\kform$ has not yet dropped enough for $\oba$ to differentiate into $\blc$. In (c), an $\oba\to\blc$ transition takes place as indicated by the arrow. The arrow starts and ends where $\kform$ is 90\% and 10\% of the drop near $\RH$ in Figure~\ref{fig:kform} (see also panel (d)). The crosses in (c) correspond to the experimental data of osteoblast surface density from~\cite{marotti-etal-1976} listed in Table~\ref{table:ob-density}, scaled to human values. (d) Profile of osteoid secretion rate $\kform(x,t)$ by active osteoblasts at $t=150\ \days$. The grey lines indicate where $\kform$ is 90\% and 10\% of the drop near $\RH$ in Figure~\ref{fig:kform} ($\oba\to\blc$ transition).}
        \label{fig:profiles}
\end{figure}

\paragraph{Evolution of the cell distribution profiles: $\bmu$ initiation and quasi-steady state}
The distribution of cells along the longitudinal~$x$ axis evolves from the initial condition into a stable multicellular travelling-wave-like structure (Figure~\ref{fig:profiles}a--c). After an initiation phase during which the shapes of the cell distribution profiles develop and stabilise, this multicellular structure progresses forward through bone without changing shape (`quasi-steady' state), with osteoclasts towards the front and osteoblasts towards the back, as in a $\bmu$ (Figure~\ref{fig:profiles}c).

Each cell type reaches a stable spatial distribution at a different rate. The population of active osteoclasts quickly increases from almost zero to its quasi-steady state within around $10~\days$. In contrast, the population of active osteoblasts takes about $130~\days$ until its spatial distribution reaches a stable shape. However, the shape of the front of the $\oba$ distribution profile (\ie, the part ahead of the maximum) takes less time to stabilise, approximately $50~\days$. The comparison of Figures~\ref{fig:profiles}b and~\ref{fig:profiles}c shows that the front of the $\oba$ distribution profile (including a part extending beyond the maximum) has reached a stable shape by $t=75~\days$. In fact, the time required for the front of the $\oba$ distribution profile to reach a steady shape ($50~\days$) corresponds to the time required to increase the population of active osteoblasts from zero to its maximum value when the $\bmu$ is initiating. All the bone traversed by the $\bmu$ at times $t\gtrsim 50~\days$, \ie\ all the bone located at $x \gtrsim -4.1~\mm$, experiences the full extent of the population of both active osteoclasts and active osteoblasts ($-4.1~\mm$ corresponds to the position of the maximum of the active osteoblast population at $t=50~\days$). The temporal pattern of osteoclast and osteoblast development seen at these fixed locations in bone whilst traversed by the $\bmu$ is no longer developing as during the initiation stage of the $\bmu$. At these locations, the $\bmu$ is seen in a quasi-steady state, and the time required to increase the population of active osteoblasts to its maximum value is reduced to about $26~\days$ (see below).

\paragraph{$\bmu$ cavity shape (closing cone)}
The $\bmu$ cavity shown at the bottom of each graph in Figure~\ref{fig:profiles}a--c progressively refills as soon as active osteoblasts are generated. During the initiation stage of the $\bmu$, the shape of the closing cone varies. When the $\bmu$ reaches its quasi-steady state, the closing cone is fully formed and progresses forward without changing shape, in association with the quasi-steady population of active osteoblasts.
Matrix apposition rate is strongly reduced when the cavity radius reaches values close to the `target' Haversian canal radius $\RH=20\ \um$. The variation of osteoid secretion rate per osteoblast along the $\bmu$, $\kform(x,t)$, is shown at $t=150\ \days$ in Figure~\ref{fig:profiles}d. The locations at which $\kform$ is 90\% and 10\% of the magnitude of the drop near $\R=\RH$ in Figure~\ref{fig:kform} (see Appendix~\ref{appx:model}) are also shown (gray lines). These locations indicate the transition of active osteoblasts ($\oba$s) into bone lining cells ($\blc$) and correspond to the start and end of the $\oba\to\blc$ transition arrow in Figure~\ref{fig:profiles}c. This transition represents the end of the (active) $\bmu$ closing cone, and the start of the (quiescent) newly-formed osteon.

\paragraph{Length of the closing cone}
Considering the start of the closing cone in Figure~\ref{fig:profiles}c to be around $x\approx -0.5~\mm$ and the end of the closing cone to correspond to the intersection of the active osteoblast surface density and bone lining cell density at around $x\approx -3~\mm$, the total length of the closing cone of the simulated $\bmu$ is about $2.5~\mm$.

\paragraph{Osteoblast density and total cell numbers in the $\bmu$}
In the steady-state $\bmu$ shown in Figure~\ref{fig:profiles}c, the population of active osteoblasts ($\oba$, blue) rises sharply at the start of the refilling process, \ie\ between $x\approx-0.35~\mm$ and $x\approx-1.12~\mm$, the latter being the position of the maximum density. This spatial interval corresponds to a time interval of about $26~\days$ $\big(\!\!\equiv\!\!\tfrac{(-1.12+0.35)~\mm}{30~\um/\da}\big)$. The peak population of active osteoblasts is reached near the disappearance of the population of pre-osteoblasts ($\obp$, cyan). Past this point, the distribution of active osteoblasts exhibits a slow, pseudo-linear decrease (from right to left) up to the point where active osteoblasts transition to bone lining cells ($\blc\to\oba$ arrow).

The dot-dashed blue line in Figure~\ref{fig:profiles}c represents the surface density of active osteoblast $\rho_\oba$ and the constant surface density of bone lining cells $\rho_\blc=2300/\mm^2$~\cite{parfitt-in-recker}. Surface densities $\rho_\oba < 2300/\mm^2$ are shaded as they do not correspond to real active osteoblasts: at these locations, osteoblasts are assumed to have become bone lining cells of constant density no longer secreting osteoid and no longer undergoing apoptosis. The transition from active osteoblasts to bone lining cells (driven by the current value of the $\bmu$ cavity radius) occurs precisely when the surface density of active osteoblasts reaches values around $2300/\mm^3$.

With the model parameters listed in Table~\ref{table:parameters} in Appendix~\ref{appx:model}, the total number of each cell type found in the simulated $\bmu$ is (up to rounding approximations):
\begin{align}
    &N_\oca \approx 100, \quad N_\ocp \approx 20, \quad N_\obu \approx 20
    \\&N_\obp \approx \num{11100}, \qquad N_\oba \approx \num{5200}.
\end{align}

The experimental data on osteoblast surface density at different cavity radii, scaled to correspond to human average $\bmu$s (crosses in Figure~\ref{fig:profiles}c), are matched by the calibrated numerical surface density $\rho_\oba$.

\begin{figure}[!tp]
    \centering
    \makebox[\figurewidth]{\input{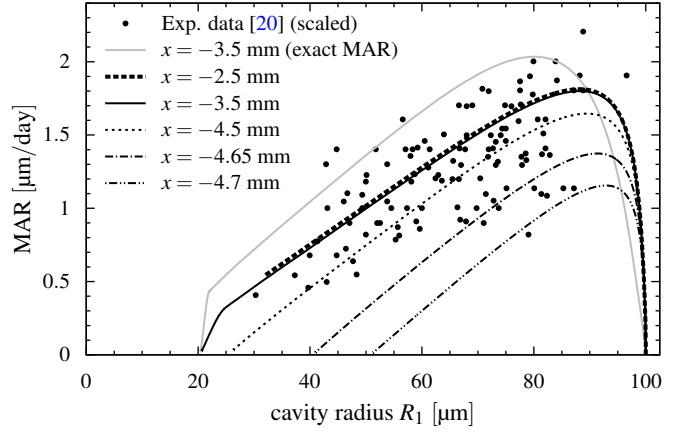}}%
    %\\\includegraphics[width=\figurewidth]{fig-mar-vs-R1-v1}
    \caption{Matrix apposition rate versus $\bmu$ cavity radius obtained from our model (lines) and from tetracycline experiments (dots). The exact matrix apposition rate $\big|\tfrac{\p}{\p t} \R(x,t)\big|$ versus $\R_1\equiv \R(x,t)$ is shown at position $x=-3.5~\mm$ in bone (solid grey line). The approximate matrix apposition rates $\big|\tfrac{\R_2-\R_1}{\Delta t}\big|\equiv\big|\tfrac{\R(x,t+\Delta t)-\R(x,t)}{\Delta t}\big|$ versus $\R_1\equiv \R(x,t)$ are shown at $x=-2.5~\mm$, $-3.5~\mm$, $-4.5~\mm$, $-4.65$, and $-4.7~\mm$ for a fixed interval $\Delta t=10~\days$. Data points are taken from Ref.~\cite[Fig.~2]{metz-etal} and have been appropriately rescaled to human average $\bmu$ sizes (see text).}
    \label{fig:mar}
\end{figure}%
\begin{figure}[!tp]
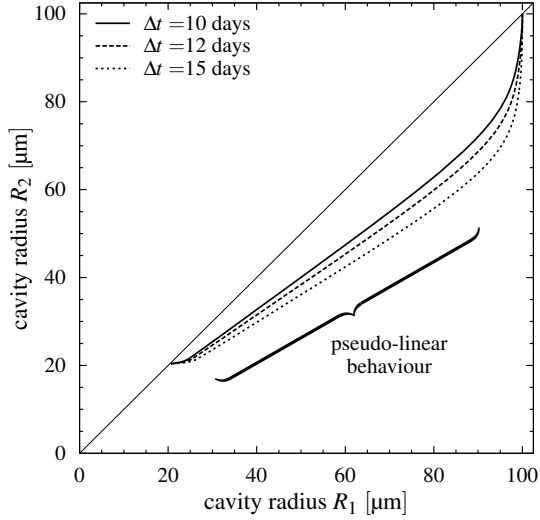

    \centering
    \makebox[\figurewidth]{\includecombinedgraphics[vecscale=0.85]{fig-R2-vs-R1}}%
    \caption{Parametric plots of cavity radius pairs $(\R_1,\R_2)$ with $\R_1=\R(x,t)$ and $\R_2=\R(x,t+\Delta t)$ obtained from our model at a fixed location in bone ($x=-4~\mm$) when $t$ evolves from $0$ to $150~\days$. The figure shows the effect of choosing different time intervals $\Delta t=$ 10 days, 12 days, and 15 days (corresponding to typical time intervals between tetracyline injections). The calculated curves intersect with the diagonal at the onset of formation and at completion of formation.}
    \label{fig:r1-r2}
\end{figure}
\paragraph{Tetracycline double labelling data}
In Figure~\ref{fig:mar} we plot the `exact', instantaneous matrix apposition rate $|\tfrac{\p}{\p t} \R(x,t)|$ versus $\R(x,t)$ at $x=-3.5~\mm$ (solid grey line) and the approximate, average matrix apposition rates ($\mar$) $\big|\tfrac{\R(x,t+\Delta t)-\R(x,t)}{\Delta t}\big|$ versus $\R(x,t)$ obtained in our simulations during refilling when $t$ evolves from $0$ to $150~\days$. The effect of sampling the radii  at various positions $x$ in bone is shown (solid and interrupted black lines), whilst the time interval $\Delta t=10~\days$ (corresponding to the interval between tetracycline injections in Ref.~\cite{metz-etal}) is kept fixed. These results are compared to experimental data from Ref.~\cite{metz-etal}. This data was collected in sheep and plotted in Ref.~\cite[Fig.~2]{metz-etal} as $\mar$ versus ``percent unfilled'', \ie, the percent of bone thickness remaining to deposit so as to reach the sheep Haversian canal radius. In Figure~\ref{fig:mar}, we rescaled this ``percent unfilled'' data to human-sized $\bmu$ cavity radii, such that 100\% unfilled corresponds to the cement line radius $\Rc=100~\um$ and 0\% unfilled corresponds to the Haversian canal radius $\RH=20~\um$ (see Eq.~\eqref{radius-scaling} in Appendix~\ref{appx:model}).

Tetracycline double labelling data is sometimes presented as a plot of the data pairs $(\R_1, \R_2)$~\cite{manson-waters,metz-etal,martin-dannucci-hood}. In Figure~\ref{fig:r1-r2} we plot the trajectory of the radii pairs $\big(\R(x,t), \R(x,t+\Delta t)\big)$ obtained during refilling for various time intervals $\Delta t$ when $t$ evolves from $0$ to $150~\days$ at the fixed location $x=-3.5~\mm$.

% \newpage
\section{Discussion}\label{sec:discussion}

\paragraph{Cell distribution profiles within the $\bmu$}
The emergence and progression of structured cell distribution profiles models the initiation and quasi-steady phases of a cortical $\bmu$. A key role is played by the biochemical coupling between the bone cells and by the cell migration properties to explain the emergence and stability of the $\bmu$~\cite{buenzli-etal-moving-bmu}. The cells present initially at $t=0$ generate a biochemical positive feedback loop, captured in the governing equations: the $\oca$s free $\tgfb$ from the bone matrix into the microenvironment, which promotes the differentiation of $\obu$s into $\obp$s. The ligand $\rankl$ expressed on $\obp$s promotes in turn the differentiation of $\ocp$s into $\oca$s, thus enabling the process to be sustained. Further towards the back of the $\bmu$, the concentration of $\tgfb$ drops~\cite{buenzli-etal-moving-bmu}. This facilitates the differentiation of $\obp$s into $\oba$s~\cite{parfitt-1994,manolagas}. At this point, bone formation starts and the cavity refills. 

The shape of the cell distribution profile of active osteoblasts ($\oba$) is appreciably modified compared to Ref.~\cite[Fig.~2]{buenzli-etal-moving-bmu} due to the additional influences on osteoblast development considered. The increase in the density of $\oba$s is sharper at the start of formation whilst its decrease towards the back of the $\bmu$ is slower, exhibiting a pseudo-linear behaviour that contrasts with the exponential-like decrease obtained in~\cite{buenzli-etal-moving-bmu}. This slower decrease is attributable to the tendency of cavity shrinkage to concentrate the density of active osteoblasts, which partly compensates the depletion pathways of the pool of active osteoblasts~\cite{pazzaglia-etal-2011}. This enabled us to calibrate the model so as to match Marotti~\etal's data on the surface density $\rho_\oba$ reasonably well.

The strong influence of cavity shrinkage on $\rho_\oba$ can be guessed by the fact that during refilling, the surface area of the $\bmu$ cavity walls shrinks to about 20\% of its initial extent (ratio $\RH/\Rc$). If the number of osteoblasts was conserved during refilling, their density would increase five-fold as they would be packed onto a much smaller surface. Osteoblast density still decreases due to their continuous elimination from the active pool, but at rate slowed down by the shrinking of the cavity.

\paragraph{Model robustness} The biochemical positive feedback loop between osteoblasts and osteoclasts is controlled (i) by elimination of the cell populations; (ii) by the negative feedback of $\tgfb$ on the population of osteoclasts (see Figure~\ref{fig:bmu-model}); and (iii) by the fact that ligand activations saturate when all the receptors on the cell membranes are occupied.
As a consequence, the qualitative behaviour of the model is generally very robust, unless drastic changes in the model structure are made (such as swapping the osteoblast cell types expressing $\rankl$ and $\opg$, or changing the cell transport properties, see \cite{buenzli-etal-moving-bmu}). The inclusion of geometrical feedbacks on the osteoblast lineage (through secretory activity and osteocyte generation) can lead in some cases to more pronounced, but still continous, changes in the density of active osteoblasts. These changes are due to the sudden drop in secretory activity when $R\approx R_H$, which is accompanied by a sudden drop in the rate of osteoblast-to-osteocyte transition (since refilling is terminating). This can lead to a temporary increase in the density of active osteoblasts when their density is high (higher than the experimental data points of \cite{marotti-etal-1976}). At normal densities, this phenomenon is manifested by a slight ``bulge'' in the density distribution near the osteoblast-to-osteocyte transition (Figure~\ref{fig:profiles}c).

\paragraph{$\bmu$ initiation and quasi-steady state}
The life history of a $\bmu$ is usually divided into three separate stages: initiation (resorption alone, early life), progression (resorption and formation, mid life), and termination (formation alone, late life)~\cite{parfitt-in-recker}. From the dynamics of cell development observed in our model, these stages can be decomposed further into several levels of quasi-steady states until the progression stage is fully developed. Indeed, whilst the population of certain cell types may have reached a quasi-steady state within the $\bmu$, the population of other cell types may still be developing. Due to the longer lifespan of osteoblasts compared to osteoclasts, we observe that osteoclast densities reach a steady state earlier in the $\bmu$'s life than osteoblast densities. The sequence of cell types reaching a steady-state spatial distribution within the $\bmu$ is: % 
(1)~precursor cells ($\obu$, $\ocp$) (this is assumed to occur instantly~\cite{buenzli-etal-moving-bmu});
(2)~active osteoclasts ($\oca$);
(3)~pre-osteoblasts ($\obp$);
(4) active osteoblasts ($\oba$).
The appearance of the first active osteoblasts in the $\bmu$ (which denotes the transition from the initiation to the progression stage according to the usual picture) occurs between stages (1) and (2). In Figure~\ref{fig:profiles}b (at $t=75~\days$), the density of active osteoclasts ($\oca$) and of pre-osteoblasts ($\obp$) have already reached a quasi-steady state distribution, but not the density of $\oba$s near the back of the $\bmu$. This state is situated between the stages (3) and (4) above, and corresponds to an early (still developing) progression stage.

According to this picture of cell and $\bmu$ development, a significant portion of the $\bmu$'s lifespan of 6--12~months~\cite{parfitt-1994} could be spent in a developing (nonsteady) stage. The stabilisation of the spatial distribution profiles of cells occurs only once the $\bmu$ has travelled sufficiently far away from its point of origin ($\approx 130~\days$). The temporal pattern of cells experienced at a single location in bone stabilises quicker (50~days). Unfortunately, this temporal pattern is not recorded in bone, unlike the spatial distribution of cells. The often depicted correspondence between temporal pattern at a fixed location and spatial distribution at a single time (see, \eg,~\cite[Figs 1--2]{parfitt-1994}) occurs only once the $\bmu$ has fully developed stable spatial distributions of all its cell types. This could have important consequences for $\bmu$-related properties of bone remodelling that are determined experimentally assuming steady-state $\bmu$s, such as resorption and formation periods and activation frequencies~\cite{martin-burr-sharkey,parfitt-in-recker}.

\paragraph{Osteoblast-to-osteocyte transition}
The generation of osteocytes occurs by embedment of active osteoblasts into the bone matrix~\cite{franzOdendaal-hall-witten,dallas-bonewald}. This complex process is believed to involve a regulation of osteoblast secretory activity by the osteocytes themselves (\eg, via sclerostin). In the model, such a regulation is not accounted for, but the rate at which the osteoblast-to-osteocyte transition needs to occur so as to reproduce a given distribution of osteocytes is calculated (see Eq.~\eqref{oba-to-ocy-rate-avg}). Since both matrix apposition rate and $\bmu$ cavity radius decrease as refilling proceeds, the rate of the osteoblast-to-osteocyte transition \eqref{oba-to-ocy-rate-avg} also decreases. A regulation by osteocytes combined with changes in the local curvature could explain this decrease~\cite{martin-2000}. However further studies are required to investigate the effect of potentially important factors, such as the mineralisation of the newly deposited matrix.

\paragraph{Osteoblast-to-bone lining cell transition}
What regulates the transition from osteoblasts to quiescent bone lining cells at the back of a $\bmu$ remains poorly understood. In the model, this transition is driven by the current $\bmu$ cavity radius (via $\kform$ in Eq.~\eqref{kform}). % and not by the current surface density $\rho_\oba$
A validation of our model of osteoblast development and of its calibration is provided by the fact that this transition occurs precisely where $\rho_\oba$ reaches observed values of bone lining cells surface densities ($\rho_\blc\approx \num{2300}/\mm^2$~\cite{parfitt-in-recker}). This is indicated in Figure~\ref{fig:profiles}c by the fact that the $\rho_\oba$ curve intersects the $\oba\to\blc$ transition arrow. When the distribution of active osteoblasts did not match the experimental points of \cite{marotti-etal-1976} in Figure~\ref{fig:profiles} due to poorly-calibrated model parameters, the following situations were encountered. For higher osteoblast densities, the cavity was refilled earlier, leading to lengths of the closing cone shorter than reported values, and to an osteoblast-to-bone lining cell transition occurring at surface densities higher than $\num{2300}/\mm^2$. For lower osteoblast densities, the $\bmu$ cavity was either (i) refilled later, leading to lengths of the closing cone larger than reported value and to an osteoblast-to-bone lining cell transition occurring at surface densities smaller than $\num{2300}/\mm^2$; or (ii) permanently under-refilled (\ie, not refilled up to the `target' Haversian canal radius $R_H$) due the population of osteoblasts becoming extinct before they could reduce their secretory activity and become lining cells.

\paragraph{Osteoblast apoptosis} Osteoblast apoptosis in a $\bmu$ is believed to serve to eliminate `surplus' osteoblasts, \ie, osteoblasts that are neither forming bone, nor becoming osteocytes, nor becoming bone lining cells~\cite{parfitt-1994}. The rate of osteoblast apoptosis, as a regulatory mechanism of osteoblast number in the closing cone, is thus likely to depend on the phase of the refilling process in the $\bmu$. As the cavity shrinks, the number of osteocytes to be generated per unit time decreases and osteoblasts are confined into a smaller volume, resulting in an increase in surplus osteoblasts. The rate of osteoblast apoptosis $A_\oba\oba$ assumed in our model, which is proportional to the population size, enabled us to obtain a profile of active osteoblast surface density that accurately matched the experimental measurements of Marotti~\etal~\cite{marotti-etal-1976} and the bone lining cell density. It is possible that comparison with more extensive data in the future may require $A_\oba$ to be dependent on the refilling stage (\eg, via $\R$).

\paragraph{Number of osteoblasts within the $\bmu$}
Precise estimates of the total number of active osteoblasts in a $\bmu$ $N_\oba$ are difficult to find in the literature. Numbers of about $3000$ per millimetre of $\bmu$ length~\cite{polig-jee}, in the range of $2000$--$4000$~\cite{jaworski-duck-sekaly}, and up to $6750$~\cite[Table~6]{parfitt-in-recker} are reported. This variation may reflect cross-sectional variability, cross-species variability, and/or measurements performed on $\bmu$s at different stages of their lifetime, and so having different closing cone lengths. Estimates of the local surface density of active osteoblasts $\rho_\oba$ are likely to be more reliable as they are based on local measurements and do not depend on the closing cone length. 

A reasonable lower bound estimate of $N_\oba$ may be found by assuming a conical shape of the closing cone between $\RH$ and $\Rc$, and a lower-bound surface density of 2300/$\mm^2$ (surface density of bone lining cells~\cite{parfitt-in-recker}). With a closing cone length $L_\form\approx 2.5~\mm$~\cite{parfitt-in-recker}, this provides the lower bound $N_\oba \gtrsim 2170$. Following Parfitt~\cite{parfitt-in-recker}, a reasonable upper-bound estimate may be found by assuming a maximum cavity--bone area given by that of a cylinder of radius $\Rc$ and an average surface density $\rho_\oba\approx 4500/\mm^2$. This provides the upper bound $N_\oba \lesssim 7000$. The total number of active osteoblasts in the steady-state $\bmu$ obtained in our simulations ($N_\oba\approx 5200$) falls well in the above range. 

The total number of pre-osteoblasts in the $\bmu$ in our simulations ($N_\obp\approx \num{11100}$) may seem high, although we are not aware of pre-osteoblast cell counts in $\bmu$s reported in the literature. The high value obtained for $N_\obp$ compared to $N_\oba$ is mainly due to the integration of $\obp$ across the whole cavity area $\Sx(x,t)$, as opposed to the integration of $\oba$ across $\SOBa(x,t)$ only. We note, however, that the fate of all pre-osteoblasts in the model is to become active osteoblasts at some point of the refilling process, \ie\ pre-osteoblasts are not assumed to undergo apoptosis. In fact, the number of pre-osteoblasts in a cross-sectional slice of thickness $\Delta x = \hOBa = 15~\um$ at a fixed position in bone reaches a maximum value of about 390 % (corresponding to a cross-sectional areal density of $2400/\mm^2$)
near the start of the refilling process, then quickly decreases to zero as all of these pre-osteoblasts become active. This activation process forms a layer of active osteoblasts against the bone surface with precisely the correct surface density, as indicated by the match of the surface density of active osteoblasts $\rho_\oba$ with the experimental data in Figure~\ref{fig:profiles}c. This shows that in terms of cell \emph{densities}, the model is well calibrated, but that \emph{absolute number of cells} derived from the model by integration of the densities may be estimated with large variabilities depending on the length of the closing cone.

\paragraph{Shape of the $\bmu$ cavity (closing cone) and Haversian canal}
The $\bmu$ cavity shown below each plot in Figure~\ref{fig:profiles}a--c is representative of the closing cone and final Haversian canal. The cutting cone was not calculated in this model because its shape depends on the movement pattern of osteoclasts within the $\bmu$~\cite{buenzli-etal-oca-resorption}.

There is not much quantitative data available on the shape of the closing cone. Longitudinal $\bmu$ sections provide an inaccurate picture of the closing cone due to section obliquity. Micro-computed tomography techniques reveal the three-dimensional shape of intracortical porosities, but are unable to distinguish quiescent surfaces (Haversian canals) from active surfaces ($\bmu$ cavities)~\cite{cooper-etal}. The length of the closing cone obtained in our simulations (around $2.5~\mm$) corresponds to values estimated for humans~\cite{parfitt-in-recker}.

The Haversian canal left after the passage of a $\bmu$ has local dimensions that depend on the stage of the $\bmu$'s life experienced at each location. The extent of both resorption and formation depends on whether osteoclasts and/or osteoblasts have reached a quasi-steady state. If quasi-steady states are reached in a sequence as our model suggests, some regions of bone may experience the full extent of resorption but not the full extent of formation. This situation occurs in our simulations near $x=-4.65~\mm$ in Figure~\ref{fig:profiles}a--c, where a mature osteoclastic front is experienced, but the initial cavity is only partially refilled because the population of osteoblasts has not fully developed. This creates a local enlargement of the Haversian canal that may be contributing to local spatial nonuniformities observed in Haversian canal sizes~\cite{cohen-harris,cooper-etal,parfitt-in-recker}. The observation of such local enlargements may be useful for the determination of a $\bmu$'s point of origin from micro-CT scans exhibiting Haversian porosities~\cite{cooper-etal}.

\paragraph{Comparison with tetracycline double labelling data}
Tetracycline label radii are usually measured on a collection of different osteons, rather than on serial sections of a single osteon~\cite{lee-1964,manson-waters,metz-etal}. Having observed great cross-sectional variations of cement line radii and Haversian canal radii, the authors of Ref.~\cite{metz-etal} have presented $\mar$ versus the percentage of unfilled bone to normalise the data from different osteons.

The refilling dynamics of our model is remarkably consistent with this data, but also emphasises some important differences (Figure~\ref{fig:mar}). For the most part, the numerical curves $\mar$ versus $\R_1$ are pseudo-linear. The continuation of the pseudo-linear segment of the solid black curve (corresponding to a steady-state $\bmu$) reaches a point near the origin, and so is in support of the linear phenomenological relationship between $\mar$ and $\R$, Eq.~\eqref{mar-vs-r}. However, the tetracycline data in Figure~\ref{fig:mar} does not reflect the dynamics of refilling of our model at the start of refilling ($\R\approx\Rc$) and at the end of refilling ($\R\approx\RH$). At these radii, the relationship between $\mar$ and $\R$ exhibited by the model is clearly nonlinear and deviates significantly from Eq.~\eqref{mar-vs-r}. In fact, Eq.~\eqref{mar-vs-r} is satified by the point $\R=0, \mar=0$, but this point is not physiologically valid since $\bmu$s stop refilling at the Haversian canal radius. In our model, the sharp increase of $\mar$ near $\Rc$ is due to the sharp rise in active osteoblast number. The sharp decrease of $\mar$ near $\RH$ is due to the secretory activity of osteoblasts quickly dropping near $\RH$.  Similar remarks apply to the numerical curves shown in Figure~\ref{fig:r1-r2}. The start and end of refilling formally correspond to points on the diagonal at $\Rc$ and $\RH$, respectively. A pseudo-linear behaviour is observed for a significant part of the numerical curves. The continuation of the pseudo-linear segment passes near the origin ($\R=0$), consistently with Eq.~\eqref{mar-vs-r} and other tetracycline data~\cite{manson-waters}. The regions of nonlinearities are associated with the sharp increase and decrease of $\mar$ at the start and end of formation. The fact that these sharp transitions in $\mar$ are not reflected in the experimental data suggests that they occur rapidly, perhaps as a synchronised event amonsts all the osteoblasts~\cite{parfitt-in-recker}. A synchronised halt of formation may allow the $\bmu$ cavity to reach a target size more consistently. One potential mechanism is that osteoblasts sense the proximity of the blood vessel or other localised components, as suggested \eg~in Ref.~\cite{qiu-parfitt-etal-2010}.

Despite data normalisation to reduce cross-sectional variability, another source of variability is manifestly present in the data from Ref.~\cite{metz-etal}. Our model suggests that the stage of the $\bmu$'s life at which the data is collected could be an important contributor to such variability. In Figure~\ref{fig:mar}, we show numerical curves obtained from regions of bone experiencing different stages of the $\bmu$'s life history. Bone near $x~\approx -4.65~\mm$ experiences the passage of the $\bmu$ in an early progression stage, with a fully-developed population of osteoclasts but a non-fully-developed population of osteoblasts. Bone at $x\gtrsim -3.5~\mm$ experiences the passage of a $\bmu$ in which all cell profiles have reached a quasi-steady state. This is indicated by the fact that the curve at $x=-2.5~\mm$ in Figure~\ref{fig:mar} overlaps with the curve at $x=-3.5 \mm$. (The curve is incomplete compared to that at $x=-3.5~\mm$ because bone located at $x=-2.5~\mm$ does not experience the passage of the back of the $\bmu$ by $t=150~\days$.) A possible explanation of the variability of the experimental data~\cite{metz-etal} is therefore that this data may have been collected from $\bmu$s at different stages of their development. The great variability observed in cement line radii is consistent with this hypothesis~\cite{metz-etal}. If so, the frequency of data points lying on curves obtained from different stages of the $\bmu$s in Figure~\ref{fig:mar} may enable an experimental determination of the relative duration of these stages, much in the same way that the frequency of $\bmu$s seen in a resorption or formation phase in histological cross sections enables the experimental determination of the duration of the resorption and formation phases~\cite{martin-burr-sharkey}.

Choosing different time intervals between the tetracycline injections has a marked effect on the slope of the pseudo-linear behaviour between $\mar$ and $\R_1$ or between $\R_2$ and $\R_1$~\cite{manson-waters}. In Figure~\ref{fig:mar}, there is an appreciable difference between the approximate matrix apposition rates $\mar=\big|\tfrac{\R(x,t+\Delta t)-\R(x,t)}{\Delta t}\big|$ calculated with $\Delta t=10~\days$ and the instantaneous matrix apposition rate $|\tfrac{\p}{\p t}\R(x,t)|$ formally corresponding to the limiting case $\Delta t\to 0$. The effect of choosing different $\Delta t$ is also shown in Figure~\ref{fig:r1-r2}. Interestingly, this choice does not particularly affect the extent of the pseudo-linear behaviour.

\paragraph{Osteoblast secretory activity}
The scarcity of experimental data on the secretory activity of individual osteoblasts in bone remodelling is perhaps surprising given that it may easily be determined by means of Eq.~\eqref{mar-vs-kform-rhooba}, combined with measurements of osteoblast surface densities and matrix apposition rates~\cite{jones,marotti-etal-1976}. The use of generalisation of Eq.~\eqref{mar-vs-kform-rhooba} to investigate tetracycline data from irregular $\bmu$s, \eg\ Eq.~\eqref{refilling-eq-surface}, may help explore the start of the refilling phase in $\bmu$, when cavity shape can be very irregular~\cite{manson-waters,rumpler-fratzl-etal}. Slightly irregular osteons were analysed in Ref.~\cite{manson-waters}, but discarded from further analysis as they did not fit Eq.~\eqref{mar-vs-r}.

In our model, osteoid secretion rate $\kform$ is determined by the current $\bmu$ cavity radius $\R$, Eq.~\eqref{kform}. The local curvature of the bone substrate may influence osteoblast membrane stress~\cite{qiu-parfitt-etal-2010}, integrin connections between neighbouring osteoblasts, and/or the transmission of osteocytic signals~\cite{marotti-2000,martin-2000}. This may influence in turn the osteoblasts' shape, size, and so their secretory activity~\cite{marotti-etal-1976,rumpler-fratzl-etal}. A biological advantage of a secretory activity effectively regulated by $\R$ is that $\bmu$ refilling becomes self-regulated: formation stops when refilling is complete. However, completion of refilling also requires that the population of active osteoblasts can be sustained long enough.

More elaborate models of osteoblast secretory activity $\kform(x,t)$ could be formulated to investigate the refilling dynamics of $\bmu$s. For example, short-term fluctuations in matrix apposition rates have been shown to occur both in animal models (24-hr to 72-hr period) and in humans (10-12 days period) \cite[and Refs cited therein]{parfitt-in-recker,martin-burr-sharkey}. Bone formation may also occasionally come to a complete rest for a long period (of the order of months), particularly with aging and in some metabolic bone diseases \cite{parfitt-in-recker}, or to exhibit work--rest cycles \cite{martin-burr-sharkey}. Osteoblast secretory activity is intimately linked with the bone matrix microstructure \cite{pazzaglia-etal-2011,pazzaglia-etal-2012-lamellar}. Some authors have proposed to link work--rest cycles or different generations of osteoblasts with the lamellar microstructure of the bone matrix. These hypotheses have been debated extensively in the literature, see \eg\  \cite{martin-burr-sharkey,parfitt-in-recker,pazzaglia-etal-2011,pazzaglia-etal-2012-lamellar}. Further insights into these questions would necessitate considerable elaborations of our computational model.

\section{Conclusions}\label{sec:conclusions}
Variations of matrix apposition rates during $\bmu$ cavity refilling were investigated in a mathematical model of cell development in a cortical $\bmu$. 

The model accounts for significant influences of the evolving cavity geometry on the density of osteoblasts. These influences moderate the decrease in osteoblast density towards the back of the $\bmu$, and enabled us to (i)~calibrate the model against experimental osteoblast densities determined at three $\bmu$ cavity radii; (ii)~retrieve matrix apposition rates decreasing linearly with $\bmu$ cavity radius as observed experimentally (Eq.~\eqref{mar-vs-r}, Figure~\ref{fig:mar}).

A significant portion of a $\bmu$'s lifespan could be spent in a nonsteady state, due to the long period required for osteoblasts to reach a steady-state distribution. This may have important repercussions for $\bmu$-related properties of bone remodelling derived under the steady-state assumption, such as resorption and formation periods, and $\bmu$ activation frequency~\cite{martin-burr-sharkey}. Osteoclast development was found to occur on a much faster timescale than osteoblast development. Portions of bone in a $\bmu$'s path may thus experience the full extent of resorption but a partial extent of formation, which could lead to local enlargements of Haversian canals near the $\bmu$'s point of origin.

The direct comparison of our model's prediction of matrix apposition rates with tetracycline data suggest that this data does not exhibit the start and the end of refilling in a $\bmu$. The build up of osteoblasts at the onset of refilling and their transition to non-synthesising bone lining cells at completion of refilling should be associated with low matrix apposition rates. The fact that this is not observed experimentally may indicate that these processes occur rapidly, as suggested by our model. Our model also suggests that a significant part of the cross-sectional variability remaining in tetracycline data after their normalisation in Ref.~\cite[Fig.~2]{metz-etal} may be explained by this data having been collected from $\bmu$s at different stages of their lifetime. 

A complete quantitative picture of the spatio-temporal dynamics of $\bmu$s based on experimental measurements remains to be elucidated. Whilst matrix apposition rates are often reported in the literature from tetracycline double labelling experiments, these data integrate both osteoblast number and osteoblast secretory activity. Furthermore, large and irregular $\bmu$s, likely to be in an early stage of refilling, are often discarded~\cite{manson-waters}. This difficulty could be alleviated by using Eq.~\eqref{refilling-eq-surface}, valid for any $\bmu$ shape, rather than Eq.~\eqref{mar-vs-kform-rhooba}, valid for cylindrical $\bmu$s. Experimental knowledge of the secretory activity of single osteoblasts is scarce, particularly in relation to a $\bmu$'s internal organisation. The determination of cell densitites and their distribution along the different phases of the $\bmu$ is essential to gain insights into the biochemical regulation of individual cells within the $\bmu$. Insights into these processes are particularly relevant for our understanding of several bone diseases, including osteoporosis~\cite{seeman}.

\subsection*{Acknowledgements}
Financial support by the Australian Research Council (Project number DP0988427 (PP, DWS) and Project number DE130101191 (PRB)) is gratefully acknowledged.

\begin{appendices}% using package appendix for more flexibility.
\section{Model description}\label{appx:model}
In this appendix, we detail some aspects of the model presented in the Methods section (Section~\ref{sec:methods}). The governing equations of all cells and signalling molecules considered in the model are presented, as well as a table listing all the model parameters.

\paragraph{Osteoblast secretory activity}
In cortical $\bmu$s, the secretory activity of single osteoblasts varies as refilling proceeds~\cite{marotti-etal-1976,volpi-etal,parfitt-in-recker,martin-burr-sharkey,zambonin-zallone-1977}. The secretion rate depends on the amount of organelles involved in glycoprotein and protein synthesis, and so is likely to depend on the cell's protoplasmic volume. The precise factors that influence osteoid secretion rate are currently not well known, but may include nutrients, hormones and regulatory molecules~\cite{parfitt-1994,manolagas,aubin-in-bilezikian}, and the local curvature of the cavity~\cite{rumpler-fratzl-etal,qiu-parfitt-etal-2010}.

To account for the variation of osteoid secretion rate during refilling in our model, we assume that $\kform$ depends on the current radius of the cavity (Eq.~\eqref{kform}): $\kform(x,t)=\kformtilde\big(\R(x,t)\big)$, with $\kformtilde(\RH)=0$, where $\RH$ is the Haversian canal radius at which osteoblasts have become (non-synthesising) bone lining cells (Figure~\ref{fig:bmu-geom}b). Marotti~\etal~\cite{marotti-etal-1976} reported osteoid secretion rates at three different radii in canine cortical $\bmu$s. We add to this data a zero secretion rate at $\RH^\text{dog}\approx 15\ \um$, the average Haversian canal radius in dogs~\cite{parfitt-in-recker} (see Table~\ref{table:kform}).
\begin{table}[!tbp]
    \centering
    \caption{Osteoid secretion rate vs $\bmu$ cavity radius (in dogs)}\label{table:kform}
    \begin{tabular}{cc}
            $\R^\text{dog}\ [\um]$ & $\kformdog\ [\um^3/\da]$
            \\\hline
            15 & $0$
            \\17& 90
            \\31 & 103
            \\65 & 180
            \\\hline
    \end{tabular}
\end{table}
\begin{figure}[!tp]
    \centering
    \makebox[\figurewidth]{\input{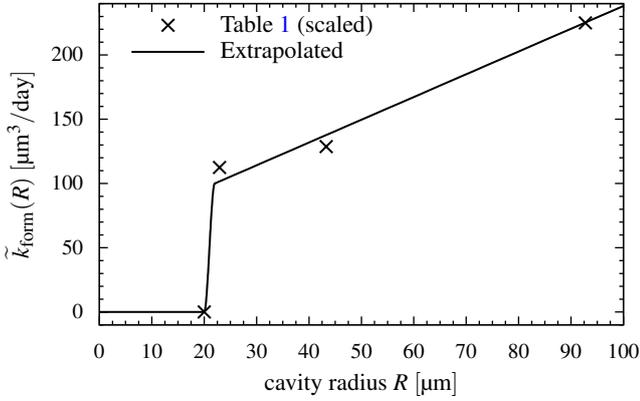}}
    \caption{Assumed functional relationship $\kformtilde(\R)$~\eqref{kform} bewteen osteoid secretion rate $\kform$ and $\bmu$ cavity radius $\R$ (solid line). Data points (crosses) are taken from Table~\ref{table:kform} and scaled according to Eqs~\eqref{radius-scaling},~\eqref{kform-scaling}.}
    \label{fig:kform}
\end{figure}
The data of Table~\ref{table:kform} is used to generate pseudo-human data by appropriate scalings as explained in the paragraph \emph{Scaling data from animal models to humans} below. The pseudo-human data is then interpolated by a continuous piecewise polynomial curve $\kformtilde(\R)$ that has a sharp stepwise increase between $\R=\RH=20\ \um$ and $\R=22~\um$, and a linear slope for $\R\geq 22~\um$, as depicted in Figure~\ref{fig:kform}. Note that all the active osteoblasts in the cross-section at $x$ secrete at the same rate $\kform(x,t)$.
As long as the cavity radius $R$ is larger than the `target' Haversian canal radius $R_H$, the activity of osteoblasts is therefore assumed continuous and uninterrupted, which is probably the most common pattern of bone refilling in cortical $\bmu$, at least in an average sense (see also comments in Section~\ref{sec:discussion}) \cite{parfitt-in-recker,martin-burr-sharkey,pazzaglia-etal-2011,pazzaglia-etal-2012-lamellar}.

\paragraph{Osteoblast-to-bone lining cell transition}
At the end of the formation phase at the back of a $\bmu$, active osteoblasts terminally differentiate into bone lining cells, thus forming a cellular layer on the newly formed bone surface~\cite{parfitt-in-recker,martin-burr-sharkey}. To account for such a transition in the model while ensuring the spatial exclusion of active osteoblasts and bone lining cells, we assume that active osteoblasts whose osteoid secretion rate $\kform$ is rapidly dropping become bone lining cells. This occurs between cavity radii $20$--$22~\um$ (Figure~\ref{fig:kform}).

\paragraph{Osteoblast-to-osteocyte transition}
A fraction of active osteoblasts is entrapped in the new bone matrix during bone formation and becomes osteocytes~\cite{marotti-2000,franzOdendaal-hall-witten,dallas-bonewald}. This is represented by the sink term $-\sigma_\ocy^\text{prod.}(x,t)$ in Eq.~\eqref{oba-balance-avg-implicit}. Here, we determine what rate of osteocyte generation $\sigma_\ocy^\text{prod.}(x,t)$ (density produced per unit time) reproduces a given osteocyte density in cortical $\bmu$s, $\ocy_\text{exp.}(x,r)$ (\eg, one determined experimentally). We allow for a possible dependence upon the radial coordinate $r$ in the cross section (Figure~\ref{fig:bmu-geom}b). 

The correspondence between $\ocy_\text{exp.}$ and $\sigma_\ocy^\text{prod.}$ is found by calculating the number of osteocytes in the slice of thickness $\Delta x$ at time $t$, $\Delta N_\ocy(x,t)$. This number increases during bone deposition from zero at the onset of bone refilling ($t=t_\text{F}$) (Figure~\ref{fig:bmu-geom}b, stage 1) to a constant final number at the end of bone refilling (Figure~\ref{fig:bmu-geom}b, stage 4). At any time $t$ in between, $\Delta N_\ocy(x,t)$ is given by (i) integrating spatially the given density $\ocy_\text{exp.}$ from $\Rc$ to $\R(x,t)$ (Figure~\ref{fig:bmu-geom}b); or (ii) integrating temporally the production rate $\sigma_\ocy^\text{prod.}$ from $t_\text{F}$ to~$t$. Equating these integrations gives:
\begin{align}
    2\pi \!\!\int_{\R(x,t)}^\Rc\hspace{-3ex} \der r\, r\, \ocy_\text{exp.}(x,r) = \int_{t_\text{F}}^t \hspace{-1ex}\der t' (\sigma_\ocy^\text{prod.}\,\SOBa)(x,t'). \notag
\end{align}
The dependence upon $t$ occurs in the integral bounds. From standard calculus, the differentiation of the above equation with respect to $t$ is given by the integrant evaluated at the time-dependent bound times the derivative of the bound (with a negative sign for the lower bound). Thus: 
\begin{align}
    -2\pi \R(x,t) &\ocy_\text{exp.}\big(x, \R(x,t)\big)\tfrac{\p}{\p t} \R(x,t) \notag
\\&= \sigma_\ocy^\text{prod.}(x,t)\,\SOBa(x,t). \notag
\end{align}
Using Eq.~\eqref{refilling-eq-radius} to substitute $\pd{}{t}\R$ and then Eq.~\eqref{oba-def} to substitute $\rho_\oba/\SOBa$, one finds:
\begin{align}\label{oba-to-ocy-rate-avg}
    \sigma_\ocy^\text{prod.}&(x,t) = 2\pi\,\R\ \ocy_\text{exp.}\,\kform\frac{\rho_\oba}{\SOBa}\notag
    \\&=\kform(x,t)\, \oba(x,t)\ \ocy_\text{exp.}\big(x,\R(x,t)\big)
\end{align}

\paragraph{Geometric influence of $\bmu$ cavity shrinkage}
The tendency of cavity shrinkage to concentrate the density of osteoblasts is represented by the source term $G\,\oba$ in Eq.~\eqref{oba-balance-avg-implicit}. To determine the form of $G(x,t)$, we consider the number of active osteoblasts in the slice at $x$ of thickness $\Delta x$, $\Delta N_\oba(x,t) = \oba\ast \SOBa\ \Delta x$. The rate of change of this number is $\tpd{}{t}\Delta N_\oba(x,t) = \left[(\tfrac{\p}{\p t}\oba) \SOBa + \oba\, \tfrac{\p}{\p t}\SOBa\right] \Delta x$. Within the slice, $\Delta N_\oba$ can only be modified by biological creation or elimination, which is described by the source and sink terms $\mathcal{D}_\obp\,\obp - A_\oba \oba - \sigma_\ocy^\text{prod.}\equiv\tpd{}{t}\oba \!-\! G\ \oba$ in Eq.~\eqref{oba-balance-avg-implicit}. Hence:
\begin{align}
    \tpd{}{t}\Delta N_\oba &= \big[(\tfrac{\p}{\p t}\oba) \SOBa + \oba\, \tfrac{\p}{\p t}\SOBa\big] \Delta x \notag
    \\&= \left[\tpd{}{t}\oba - G\ \oba\right] \SOBa\ \Delta x.
\end{align}
Dividing by $\oba\ast\SOBa\ \Delta x$, this gives:
\begin{align}\label{G}
    G(x,t) &= - \frac{\tfrac{\p}{\p t} \SOBa(x,t)}{\SOBa(x,t)} = - \frac{\tfrac{\p}{\p t} \R(x,t)}{\R(x,t)\Big(1-\frac{\hOBa}{2\R(x,t)}\Big)}\notag
\\&=-\kform(x,t)\,\oba(x,t)\,\frac{\hOBa}{\R(x,t)},
\end{align}
where Eqs~\eqref{soba} and~\eqref{refilling-eq-radius-oba} were used for the last equalities.

\paragraph{Scaling data from animal models to humans}
To calibrate our model to a human cortical $\bmu$, we scale $\bmu$-related data obtained from animal models according to known (or suspected) cross-species differences.

The $\bmu$ cavity radius data $\R^\text{dog}$ from Table~\ref{table:kform} was scaled to (\mbox{pseudo-})human values $\R$ by assuming a linear relationship $\R=\R(\R^\text{dog})$ such that $\R(\RH^\text{dog}) = \RH$ and $\R(\Rc^\text{dog}) = \Rc$. This gives:
\begin{align}\label{radius-scaling}
    \R&(\R^\text{dog}) = \RH + \frac{\Rc-\RH}{\Rc^\text{dog}-\RH^\text{dog}}(\R^\text{dog}-\RH^\text{dog}).
\end{align}
We used the typical values  $\RH^\text{dog} \approx 15~\um$ and  $\RH\approx 20~\um$ for canine and human Haversian canal radii. In Ref.~\cite{parfitt-in-recker}, the average cement line radius in dogs is $\Rc^\text{dog}\approx 60~\um$. In Ref.~\cite{jowsey}, it is $\Rc^\text{dog}\approx 77~\um$. Since one of the three $\bmu$ cavities in Ref.~\cite{marotti-etal-1976} has a radius $65~\um$ in the formation phase, we take here $70~\um$ as the representative canine cement line radius. We used the value $\Rc\approx 100~\um$ for the human cement line radius~\cite{jowsey,parfitt-in-recker,martin-burr-sharkey}.

The osteoid secretion rates $\kformdog$ of Table~\ref{table:kform} estimated by Marotti~\etal~\cite{marotti-etal-1976} were scaled by a factor $\alpha_\kform=1.25$ to account for higher secretion rates in humans~\cite{polig-jee}, \ie:
\begin{align}\label{kform-scaling}
    \kform = \alpha_\kform \kformdog.
\end{align}
The model of Polig and Jee~\cite{polig-jee} suggests that $\alpha_\kform$ may take values from 1.27 to 1.7. Here, the specific numerical value of the factor $\alpha_\kform$ is chosen so as to reduce the total number of active osteoblasts in the $\bmu$ (see Results and Discussion, Sections~\ref{sec:results},\ref{sec:discussion}). The scaled radius and osteoid secretion rate data of Table~\ref{table:kform} are shown in Figure~\ref{fig:kform} (crosses).

Scalings are also applied to the data of Table~\ref{table:ob-density} reporting canine osteoblast surface densities $\rho_\oba^\text{dog}$ at three $\bmu$ cavity radii. The radii $\R^\text{dog}$ are scaled to correspond to human average $\bmu$s as in Eq.~\eqref{radius-scaling}. The osteoblast surface densities $\rho_\oba^\text{dog}$ are scaled to pseudo-human data as $\rho_\oba = \alpha_\kform^{-1}\rho_\oba^\text{dog}$. The choice of the scaling factor $\alpha_\kform^{-1}=0.8$ is motivated by the corresponding scaling factor $\alpha_\kform=1.25$ performed on $\kformdog$. Indeed, increased osteoid secretion rate is associated with increased cell volume, and thus with a corresponding decrease in cell density~\cite{volpi-etal,marotti-etal-1976}.

\begin{table}[!tp]
    \centering%
    \caption{Osteoblast surface density vs $\bmu$ cavity radius (in dogs)}\label{table:ob-density}
    \begin{tabular}{cc}
            $\R^\text{dog}\ [\um]$ & $\rho_\oba^\text{dog}\ [\mm^{-2}]$
            \\\hline
            17& 4500
            \\31 & 8770
            \\65 & 10000
            \\\hline
    \end{tabular}
\end{table}

\paragraph{Scaling of the cell density distributions in the $\bmu$}
The cell distribution profiles along $x$ resulting from the numerical simulation are calibrated against experimental values of osteoblast surface densities and total cell numbers. To this end, we have modified the values of a number of parameters compared to the parameters used in Ref.~\cite{buenzli-etal-moving-bmu}. We used a scaling scheme of the model parameters such that $\oba$, $\obp$, $\obu$, $\oca$ and $\obp$ would scale uniformly without affecting much their own spatial distribution and without affecting their spatial relation with other cell density profiles. We defined scaling factors $\alpha_\oba$, $\alpha_\obp$, $\alpha_\obu$, $\alpha_\oca$ and $\alpha_\ocp$, such that if $\alpha_\oba=500$, the density of $\oba$s is multiplied by a factor 500 etc. Based on Eqs~\eqref{governing-eqs-pdes}--\eqref{piact-pirep}, the new parameter values (denoted below by a prime) ensuring such scaled densities were determined by multiplying the previous parameter values (non-primed) with an adequate combination of the above scaling factors, according to:
\begin{align}
    &\obumax' = \alpha_\obu\ \obumax, &\quad&\ocpmax' = \alpha_\ocp\ \ocpmax, \notag
    \\&D_\obu' = \frac{\alpha_\obp}{\alpha_\obu} D_\obu, &\quad &D_\ocp' = \frac{\alpha_\oca}{\alpha_\ocp} D_\ocp, \notag
    \\&n_{\tgfb}^{\text{bone}'} = n_\tgfb^\text{bone}/\alpha_\oca, &\quad &N^{\rank '}_\ocp = N^\rank_\ocp / \alpha_\ocp, \notag
    \\&\beta^{\rankl'}_\obp = \beta^\rankl_\obp / \alpha_\obp, &\quad &\beta^{\opg '}_{\oba} = \beta^{\opg}_{\oba} / \alpha{_\oba},\label{scaling}
\end{align}
where $\obumax$ and $\ocpmax$ denote parameters scaling the given distributions of $\obu$s and $\ocp$s~\cite{buenzli-etal-moving-bmu}. We note that it is not possible to control the scaling of $\obp$ independently from that of $\oba$ without affecting their spatial relationship, and so we have always taken $\alpha_\obp = \alpha_\oba$ when calibrating the model.

\paragraph{Governing equations of cell densities and signalling molecules concentrations}
Below we summarise all the governing equations of the model. The material balance equation governing the dynamics of active osteoblasts, Eq.~\eqref{oba-balance-avg-implicit}, is rewritten using the explicit expressions in Eqs \eqref{oba-to-ocy-rate-avg} and~\eqref{G}:
\begin{align}
    \tfrac{\p}{\p t} \oca = &\docp(\rankl)\,\ocp - \aoca(\tgfb)\,\oca - \tfrac{\p}{\p x} \big(\oca v_\oca\big),\notag%\label{oca-balance}
    \\\tfrac{\p}{\p t} \obp =& \dobu(\tgfb)\,\obu - \dobp(\tgfb)\,\obp,\notag%\label{obp-balance}
    \\\tfrac{\p}{\p t}\oba=&\mathcal{D}_\obp(\tgfb) \obp - A_\oba \oba \notag%\label{oba-balance-avg-explicit}
    \\ &+ \kform\, \oba\Big[\!\tfrac{\hOBa}{\R}\oba \!-\! \ocy_\text{exp.}\big(x,\R\big)\!\!\Big],\notag
    \\\tfrac{\p}{\p t} \tgfb =& n_\tgfb^\text{bone}\,\kres\, \oca - D_\tgfb\,\tgfb,\label{governing-eqs-pdes}
\end{align}
where
\begin{align}
    &\docp(\rankl) = D_\ocp\piact\big(\tfrac{\rankl}{k^\rankl_\ocp}\big),\notag
    \\&\aoca(\tgfb) = A_\oca\piact\big(\tfrac{\tgfb}{k^\tgfb_\oca}\big),\notag
    \\&\dobu(\tgfb) = D_\obu\piact\big(\tfrac{\tgfb}{k^\tgfb_\obu}\big),\notag
    \\&\dobp(\tgfb) = D_\obp\pirep\big(\tfrac{\tgfb}{k^\tgfb_\obp}\big),\label{governing-eqs-rates}
\end{align}
and
\begin{align}
    \pth(x,t) = &\beta_\pth / D_\pth, \notag% \label{pth}
    \\\rank(x,t) = &N^\rank_\ocp\, \ocp, \notag
    \\\opg(x,t) = &\frac{\beta^\opg_\oba\, \oba\, \pirep\!\Big(\frac{\pth}{k^\pth_{\ob,\text{rep}}}\Big)}{\beta^\opg_\oba\,\oba\,\pirep\!\Big(\frac{\pth}{k^\pth_{\ob,\text{rep}}}\Big)/\opg_\text{sat} + D_\opg} \notag
    \\\rankl(x,t) = &\frac{\beta^\rankl_\obp\,\obp}{1+k^\rankl_\rank\,\rank + k^\rankl_\opg\,\opg} \notag
    \\&\times \left\{\rule{0pt}{4ex}\right. \!\!D_\rankl + \frac{\beta^\rankl_\obp\,\obp}{N^\rankl_\obp\obp\ \piact\Big(\frac{\pth}{k^\pth_{\ob,\text{act}}}\Big)}\!\!\left. \rule{0pt}{4ex}\right\}^{-1}.\label{governing-eqs-aes}
\end{align}
The density distributions $\obu(x,t)$ and $\ocp(x,t)$ are assumed to be given functions (see Table~\ref{table:parameters}). The dimensionless functions $\piact$ and $\pirep$ express the activation and repression of cell differentiation, apoptosis, or ligand expression by regulatory signalling molecules. These functions are determined by the fraction of receptors on the cell occupied by the concerned signalling molecules (see Refs~\cite{pivonka-etal1,buenzli-etal-moving-bmu}). Mathematically, they are so-called Hill functions given by:
\begin{align}\label{piact-pirep}
    \piact(\xi) = \frac{\xi}{1+\xi}, \qquad \pirep(\xi) = 1-\piact(\xi) = \frac{1}{1+\xi}.
\end{align}

A slight change in the expression for $\rankl$ in Eq.~\eqref{governing-eqs-aes} has been made compared to Ref.~\cite{buenzli-etal-moving-bmu}. The production of $\rankl$ is now correctly proportional to the number of cells that express $\rankl$, \ie, we have replaced $\beta_\rankl$ in Ref.~\cite[Eq.~(19)]{buenzli-etal-moving-bmu} by $\beta_\obp^\rankl\,\obp$ in Eq.~\eqref{governing-eqs-aes}. The same inconsistency of having a production rate of $\rankl$ not scaled by the number of osteoblasts is present in the temporal model of Ref.~\cite{pivonka-etal1}, which was corrected in Ref.~\cite{buenzli-etal-anabolic}. The behaviour of the $\bmu$ model is not changed significantly by this correction. Some inconsistent behaviours of the temporal model~\cite{pivonka-etal1} were corrected by this change, see Ref.~\cite{buenzli-etal-anabolic} for more details.

\paragraph{$\bmu$ model parameters and functions}
All the parameters used in the model are listed in Table~\ref{table:parameters}. Several parameters were taken over from the temporal model of bone remodelling of Refs~\cite{pivonka-etal1,buenzli-etal-anabolic} and our previous model of the $\bmu$~\cite{buenzli-etal-moving-bmu} (see last column of Table~\ref{table:parameters}). We emphasise that rate parameters for cell differentiation/apoptosis are weighted by Hill functions of the concentration of signalling molecules, Eqs~\eqref{governing-eqs-rates}, which vary themselves in space and time. These rate parameters were therefore calibrated to known $\bmu$ characteristics that they determine, and should not be viewed as the actual value of the differentiation/apoptosis rates. For example, $A_\oca$ was chosen such that the $\oca$ population reaches a steady state within 10 days (corresponding to the lifespan of osteoclast nuclei in active osteoclasts~\cite{jaworski-duck-sekaly}), but $1/A_\oca \neq 10~\days$.

\begin{table*}[!p]
    \centering
        \vspace{-3mm}
    \caption{$\bmu$ model parameters and given functions}\label{table:parameters}
  \small
        %% \begin{tabularx}{\textwidth}{@{}lrX@{}}%@{} suppresses intercolumn space. X is an extensible justified column.
        \begin{tabular}{l@{}rp{0.36\textwidth}p{0.3\textwidth}@{}}
        \toprule
        Symbol & Value & Description & Origin of the estimate/References
        \\\thickmidrule
        $\ocp(x,t)$ & (given function) & pre-osteoclast density: gaussian function centred at $x(t)=-0.35~\mm+ v_\oca t$, with standard deviation $=0.1~\mm$ and amplitude $=2589/\mm^3$. & Standard deviation set as half the reversal zone length estimated at 200~\um~\cite{martin-burr-sharkey,parfitt-1994}; amplitude calibrated for $N_\ocp$.\bstrut{1.5ex}
        \\$\obu(x,t)$ & (given function) & uncommitted osteoblast progenitors (\msc) density, equal to $\ocp(x,t)$ & Same standard deviation as $\ocp(x,t)$; amplitude calibrated for $N_\obu$.\bstrut{1.5ex}
        \\$\kformtilde(\R)$ & (given function) & volume of osteoid secreted per osteoblast per day & Extrapolated from \cite{marotti-etal-1976} (see Eq.~\eqref{kform} and Figure~\ref{fig:kform})
        \\\midrule
        $D_\ocp$ & $41.26/\da$& $\ocp\to\oca$ differentiation rate parameter &Calibration for $N_\ocp$, $N_\oca$, Eqs~\eqref{scaling}\bstrut{1.5ex}
        \\$A_\oca$ & $2.82/\da$& \oca\ apoptosis rate parameter & \cite{buenzli-etal-moving-bmu} (calibrated for $\oca$s to reach steady state within 10 days ($\approx$ lifespan of osteoclast nuclei in active osteoclasts~\cite{jaworski-duck-sekaly}))\bstrut{1.5ex}
        \\$D_\obu$ & $81.07/\da$ & $\obu\to\obp$ differentiation rate parameter & Calibration for $N_\obu,N_\oba$, Eqs~\eqref{scaling}
        \\$D_\obp$ & $0.166/\da$& $\obp\to\oba$ differentiation rate parameter &\cite{buenzli-etal-moving-bmu,buenzli-etal-anabolic}
        \\$A_\oba$ & $0.0385/\da$& \oba\ apoptosis rate & Calibration for $\rho_\oba$ distribution profile
        \\\midrule
        \tablestrut$k^\rankl_\ocp$ &$\num{1.0025e7} \mm^{-3}$& parameter for \rankl\ binding on \ocp & \cite{buenzli-etal-moving-bmu,buenzli-etal-anabolic}
        \\\tablestrut$k^\tgfb_\oca$ &$\num{339.2}\ \mm^{-3}$& parameter for \tgfb\ binding on \oca & \cite{buenzli-etal-moving-bmu,buenzli-etal-anabolic}
        \\\tablestrut$k^\tgfb_\obu$ &$\num{339.2}\ \mm^{-3}$& parameter for \tgfb\ binding on \obu &\cite{buenzli-etal-moving-bmu,buenzli-etal-anabolic}
        \\\tablestrut$k^\tgfb_\obp$ &$\num{105.6}\ \mm^{-3}$& parameter for \tgfb\ binding on \obp &\cite{buenzli-etal-moving-bmu,buenzli-etal-anabolic}
        \\\tablestrut$k^\pth_{\ob,\text{act}}$ &$\num{9.033e7}\ \mm^{-3}$& parameter for \pth\ binding on \ob\ (for $\piact$) &\cite{buenzli-etal-moving-bmu,buenzli-etal-anabolic}
        \\\tablestrut$k^\pth_{\ob,\text{rep}}$ &$\num{134039}\ \mm^{-3}$& parameter for \pth\ binding on \ob\ (for $\pirep$) &\cite{buenzli-etal-moving-bmu,buenzli-etal-anabolic}
        \\\tablestrut$k^\rankl_\rank$ &$\num{5.6655e-8}\ \mm^3$& association binding constant for \rankl\ and \rank &\cite{buenzli-etal-moving-bmu,buenzli-etal-anabolic}
        \\\tablestrut$k^\rankl_\opg$ &$\num{1.66058e-9}\ \mm^3$& association binding constant for \rankl\ and \opg &\cite{buenzli-etal-moving-bmu,buenzli-etal-anabolic}
        \\$\beta^\rankl_\obp$ &$\num{348.377}/\da$& production rate of \rankl\ per \obp & Calibration for $N_\oba$, Eqs~\eqref{scaling}
        \\\tablestrut$\beta^\opg_\oba$ &$\num{326305}/\da$& production rate of \opg\ per \oba & Calibration for $N_\oba$, Eqs~\eqref{scaling}
        \\\tablestrut$\beta_\pth$ &$\num{1.506e8}\ \mm^{-3}/\da$& production rate of systemic \pth &\cite{buenzli-etal-moving-bmu,buenzli-etal-anabolic}
        \\\tablestrut$N^\rankl_\obp$ & \num{2.7e6}& maximum number of \rankl\ per \obp &\cite{buenzli-etal-moving-bmu,buenzli-etal-anabolic}Lemaire=\num{3e6}
        \\\tablestrut$N^\rank_\ocp$ & \num{2326}& number of \rank\ receptors per \ocp & Calibration for $N_\ocp$, Eqs~\eqref{scaling}
        \\$\opg_\text{sat}$ & $\num{1.205e14}\ \mm^{-3}$ & \opg\ density at which $\opg$ production stops &\cite{buenzli-etal-moving-bmu,buenzli-etal-anabolic}
        \\$D_\tgfb$ &$0.5/\da$& degradation rate of \tgfb &\cite{buenzli-etal-moving-bmu} (reduced compared to~\cite{buenzli-etal-anabolic} to broaden the distribution of $\tgfb$ within the $\bmu$)
        \\$D_\rankl$ &$10.13/\da$& degradation rate of \rankl &\cite{buenzli-etal-moving-bmu,buenzli-etal-anabolic}Lemaire
        \\$D_\opg$ &$0.35/\da$& degradation rate of \opg &\cite{buenzli-etal-moving-bmu,buenzli-etal-anabolic}
        \\$D_\pth$ &$86/\da$& degradation rate of \pth &\cite{buenzli-etal-moving-bmu,buenzli-etal-anabolic}
        \\\midrule
        $n_\tgfb^\text{bone}$ & $\num{3946}~\mm^{-3}$ & concentration of \tgfb\ stored in the bone matrix & Calibrated for $N_\oca$, Eqs~\eqref{scaling}
        %% \\\midrule
        \\$\kres$ & $\num{9.425e-6}~\mm^3/\da$ & volume of bone resorbed per osteoclast per day &set to $\pi \Rc^2 v_\oca/N_\oca$
        \\\midrule
        $v_\oca$ & 30~\um/\da & speed of active osteoclasts and of the \bmu & \cite{martin-burr-sharkey,parfitt-1994,parfitt-in-recker}
        \\$\RH$ & 20~\um & human Haversian canal radius (radius at which osteoblast secretory activity has stopped)&\cite{jowsey,parfitt-in-recker,martin-burr-sharkey}
        \\$\Rc$ & 100~\um & human cement line radius (taken as initial cavity radius) &\cite{jowsey,parfitt-in-recker,martin-burr-sharkey}
        \\$\hOBa$ & 15~\um & average height of an active osteoblast &\cite{martin-burr-sharkey,zambonin-zallone-1977}
        \\$\alpha_\kform$ & 1.25 & canine-to-human conversion factor for osteoid secretion rate & Calibration for $N_\oba$ (see Appendix~\ref{appx:model})
        \\$\rho_\blc$ & $\num{2300}/\mm^2$ & surface density of bone lining cells &\cite{parfitt-in-recker}
        \\$\ocy_\text{exp.}(x,r)$ & $\num{20000}/\mm^3$ & volumetric density of osteocytes in the $\bmu$ &\cite{parfitt-in-recker}
        \\\bottomrule
    \end{tabular}
    %% \end{tabularx}
\end{table*}

\end{appendices}

%\clearpage % flush table before references.

\end{document}